\pdfoutput=1

\documentclass[conference]{IEEEtran}
\usepackage{svg}
\usepackage{listings}
\usepackage[font=small]{caption}
\usepackage{array}
\usepackage{graphicx}
\usepackage{courier}
\usepackage{xcolor}
\usepackage{amsfonts}
\usepackage[utf8]{inputenc}
\usepackage{enumitem}

\graphicspath{ {./figures/} }
\newcommand{\sys}{Sharingan}
\newcommand{\cir}[1]{(#1)}

\definecolor{codegreen}{rgb}{0,0.6,0}
\definecolor{codegray}{rgb}{0.5,0.5,0.5}
\definecolor{codepurple}{rgb}{0.58,0,0.82}
\definecolor{backcolour}{rgb}{0.86,0.86,0.85}

\lstdefinestyle{mystyle}{
    backgroundcolor=\color{backcolour},
    commentstyle=\color{codegreen},
    keywordstyle=\color{magenta},
    numberstyle=\tiny\color{codegray},
    stringstyle=\color{codepurple},
    basicstyle=\ttfamily\footnotesize,
    breakatwhitespace=false,
    breaklines=true,
    captionpos=b,
    keepspaces=true,
    numbers=left,
    numbersep=5pt,
    showspaces=false,
    showstringspaces=false,
    showtabs=false,
    tabsize=2
}

\usepackage{pifont}
\newcommand{\cmark}{\textcolor{green}{\ding{51}}}%
\newcommand{\xmark}{\textcolor{red}{\ding{55}}}%

\ifCLASSINFOpdf
\else
\fi

\begin{document}
%
\title{Session-layer Attack Traffic Classification by Program Synthesis}



%

\author{\IEEEauthorblockN{Lei Shi\IEEEauthorrefmark{1},
Yahui Li\IEEEauthorrefmark{2},
Rajeev Alur\IEEEauthorrefmark{1}, 
Boon Thau Loo\IEEEauthorrefmark{1}
}
\IEEEauthorblockA{\IEEEauthorrefmark{1}University of Pennsylvania, Philadelphia PA 19104, USA\\
\{shilei,boonloo,alur\}@seas.upenn.edu}
\IEEEauthorblockA{\IEEEauthorrefmark{2}Tsinghua University, Beijing, China\\
li-yh15@mails.tsinghua.edu.cn}}


\IEEEoverridecommandlockouts
\makeatletter\def\@IEEEpubidpullup{6.5\baselineskip}\makeatother
\IEEEpubid{\parbox{\columnwidth}{
}
\hspace{\columnsep}\makebox[\columnwidth]{}}

\maketitle

\begin{abstract}
Writing classification rules to identify malicious network traffic is a
time-consuming and error-prone task. Learning-based classification
systems automatically extract such rules from positive and negative
traffic examples. However, due to limitations in the representation of
network traffic and the learning strategy, these systems lack both
expressiveness to cover a range of attacks and interpretability in fully
describing the attack traffic’s structure at the session layer. This
paper presents Sharingan system, which uses program synthesis techniques
to generate network classification programs at the session layer.
Sharingan accepts raw network traces as inputs, and reports potential
patterns of the attack traffic in NetQRE, a domain specific language
designed for specifying session-layer quantitative properties. Using
Sharingan, network operators can  better analyze the attack pattern due
to the following advantages of Sharingan’s learning process: (1) it
requires minimal feature engineering, (2) it is amenable to efficient
implementation of the learnt classifier, and (3) the synthesized program
is easy to decipher and edit. We develop a range of novel optimizations
that reduce the synthesis time for large and complex tasks to a matter
of minutes. Our experiments show that Sharingan is able to correctly
identify attacks from a diverse set of network attack traces and
generates explainable outputs, while achieving accuracy comparable to
state-of-the-art learning-based intrusion detection systems.
\end{abstract}


%

\section{Introduction}
\label{sec:intro}
Network monitoring and intrusion detection systems are essential for network 
infrastructure management. These systems require classification of network 
traffic at their core. Today, network operators and equipment vendors write 
classification programs or patterns upfront in order to differentiate 
malicious flows from normal ones. The process of writing these classification 
programs often requires deep operator insights, can be error prone, and is 
not easy to extend to handle new scenarios. 

There have been many attempts at automated learning-based classifiers for 
malicious traffic using machine learning\cite{mirsky2018kitsune, zhang2015robust, berezinski2015entropy, kwon2017survey} 
and data mining\cite{bortolameotti2017decanter,singh2004automated,wang2012generating,zhang2014toward,newsome2005polygraph} techniques. 
These classifiers have not gained traction in production systems, in part due 
to unavoidable false positive reports and the gap between the learning output 
and explainable operational insights\cite{sommer2010outside}. The challenges call for a more
expressive, interpretable and maintainable learning-based classification system.

Specifically, existing approaches suffer from the following limitations. 
First, regular expression and feature vector representations frequently 
used for input data lack session-layer details and intermediate states 
in network protocols. Hence, information about vulnerabilities in application-layer 
protocols or multi-stage attacks will be missed. Second, these 
representations also require laborious task-specific feature engineering 
to get effective learning results, which undermines the systems' advantages 
of automation. Third, it is hard to interpret the learning results 
to understand the intent and structure of the malicious traffic, 
due to the blackbox model of many machine-learning approaches and the 
lack of expressiveness in the inputs and outputs to these learning systems.


To address the above limitations, we introduce \sys{}, which uses program synthesis 
techniques to auto-generate network classification programs from labeled examples of
network traffic traces.  \sys{} aims to bridge the gap between learning systems and 
operator insights, by identifying properties of the attack that can help inform 
the network operators on the nature of the attack, and provide a basis for automated 
generation of the classification rules. \sys{} does not aim to outperform state-of-the-art 
learning systems in accuracy, but rather match their accuracy, while generating output 
that is more explainable and easier to maintain.

To achieve these goals, we adopt techniques from {\em syntax guided program synthesis} \cite{alur2013syntax} 
to generate a NetQRE~\cite{netqre} program that distinguishes the positive 
and negative examples. NetQRE, which stands for Network {\em Quantitative Regular Expressions}, 
enables quantitative queries for network traffic, 
based on flow-level regular pattern matching. The classification is done by comparing 
the synthesized program's output for each example with a learnt threshold $T$. 
Positive examples fall above $T$. The synthesized NetQRE program serves the role of network 
classifier, identifying flows which match the program specifications.

\sys{} has the following key advantages over prior approaches, which either rely on keyword and 
regular expression generation~\cite{bortolameotti2017decanter,singh2004automated,wang2012generating,zhang2014toward,newsome2005polygraph} 
or statistical anomaly detection~\cite{mirsky2018kitsune, zhang2015robust, berezinski2015entropy, kwon2017survey}.

\noindent {\bf Requires minimal feature engineering:} NetQRE~\cite{netqre} is an expressive language, and
allows succinct description of a wide range of tasks ranging from detecting security
attacks to enforcing application-layer network management policies.
\sys{} can synthesize any network task on raw traffic expressible as a NetQRE program, 
without any additional feature engineering. This is an improvement over systems based
on manually extracted feature vectors.


\noindent {\bf Efficient implementation: } The NetQRE program synthesized by \sys{} 
can be compiled, as has been shown in prior work~\cite{netqre}, to efficient low-level 
implementations that can be integrated into routers and other network devices.  
On the other hand, traditional statistical 
classifiers are not directly usable or executable in network filtering systems.  

\noindent {\bf Easy to decipher and edit:} Finally, \sys{} generates NetQRE programs 
that can be read and edited. Since they are generic executable programs with
high expressiveness, the patterns in the program reveal the stateful protocol 
structure that is used for the classification, which blackbox statistical models,
packet-level regular expressions and feature vectors have difficulty describing. 
The programs are also amenable to calibration 
by a network operator, for example, to mix in local policies or debug.


The key technical challenge in design and implementation of \sys{} is the computationally
demanding problem of finding a NetQRE expression that is able to separate positive network 
traffic examples from the negative ones.
This search problem is an instance of the so-called {\em syntax-guided synthesis} \cite{alur2013syntax}.
While this problem has received a lot of attention in recent years, no existing
tools and techniques can solve the instances of interest in our context due to 
the complexity of the expressions to be synthesized as well as the scale of 
the data set of network traffic examples used in training.
To address this challenge, we devised two novel techniques for optimizing the search --
\textit{partial execution} and \textit{merge search}, which effectively achieve 
orders of magnitude reduction in synthesis time. We summarize our key contributions:

\noindent {\bf Synthesis-based classification architecture.} We propose the methodology of reducing network attack traffic classification problem to a synthesis from examples. 

\noindent {\bf Efficient synthesis algorithm } We devise two efficient algorithms: 
\textit{partial execution}
and \textit{merge search}
, which efficiently explore the program space and enable learning from very large data sets. 
Independent of our network traffic classification use cases, 
these algorithms advance the state-of-the-art in program synthesis.

\noindent{\bf Implementation and evaluation. } We have implemented \sys{} 
and evaluated it for a rich set of metrics using the CICIDS2017~\cite{sharafaldin2018toward} 
intrusion detection benchmark database. \sys{} is able to synthesize a large range of 
network attack classification programs in a matter of minutes with accuracy comparable to
state-of-the-art systems. Moreover, the generated NetQRE program is easy to interpret, 
tune, and can be compiled into configurations usable by existing network filtering systems.

\section{Background}
\label{sec:background}

In this section, we provide background knowledge to motivate our work and help readers put in context our techniques later in the paper.

\subsection{Learning-based Systems}
Learning-based systems face extra difficulties in network security applications compared
to traditional use cases such as recommendation systems, spam mail filtering or OCR~\cite{sommer2010outside}.
It is mainly due to two reasons. First, misclassifications in network attack detection 
has tangible cost. For example, in intrusion detection systems,
operators will need to manually investigate possible intrusion reports in order to take
proper mitigation actions. False positive reports will cost extra human labor for the
investigation. Reports in a format hard to interpret will make the process even more difficult.
Second, networks in production environments typically sends large volumes of data,
which amplifies the misclassification costs. Making things worse, network environments have 
high diversity, and are in constant flux. Ensuring absolute accuracy at training time 
is difficult. The learnt classifiers need the ability to be analyzed and maintained.

Many machine learning based systems inherently have usability issue~\cite{sommer2010outside}, 
since they do not give
interpretable reports and are hard to dynamically update. Even for those machine learning
systems with interpretable models such as decision trees~\cite{yang2019tree}, or signature learning 
systems based on data mining 
techniques~\cite{bortolameotti2017decanter,singh2004automated,wang2012generating,zhang2014toward,newsome2005polygraph}, 
the lack of expressiveness in the inputs and outputs 
to these tools undermines the efficiency of the process. 
We elaborate on these challenges next.

\subsection{Requirements for Trace Analysis}
The input representation and output model determines the
capability and usability of a learning-based system. A good model choice should 
address the following concerns to fully capture the information embedded in
a network trace. First, a network trace is a sequence of packets of varying lengths listed 
in increasing timestamp order. 
As an element of the sequence, each packet is a list of header fields.
The input representation should not discard the rich information contained in this
potentially large amount of data, and at the same time be adaptive to the varying length.
Second, network protocols are state machines by design. The output model should
be able to express the change of internal states of the flows.

No existing learning-based system can meet the above requirements in an efficient manner. 
Broadly speaking, their input/output formats 
fall into the following categories:
(1) feature vector to statistical relations between features~\cite{mirsky2018kitsune, zhang2015robust, berezinski2015entropy, kwon2017survey}, 
(2) feature vector to heuristics-based relations between features
~\cite{bortolameotti2017decanter,singh2004automated,zhang2014toward,newsome2005polygraph}, 
(3) payload string to regular expression patterns~\cite{wang2012generating}, 
and (4) network trace to state machine patterns~\cite{luo2013position, moon2019alembic}.

The categories (1) and (2) above characterize the majority of systems using machine learning or data mining
techniques. The input features are typically statistical values at the entire
flow level, such as the total duration of a flow, mean forward packet length, 
min activation time, etc. These systems have difficulty capturing the patterns
of attacks that require tracking the change of protocol states.  For example,
to learn the Slowloris attack, one needs to recognize the establishment of a large
number of handshakes, and at the same time an excessively low transmission rate or long
duration in each established session. This combination is beyond the expressiveness
of simple flow statistics.

Although one can theoretically use more complex features such as 
'number of established handshakes', 'average transmission rate' and 
'average flow duration' to learn this attack,
there are two problems in doing so. In terms of input to the learning system, 
the discovery of these
features are manually done by the user rather than automatically extracted 
from the traces.  In terms of output of the learning system, even if learning-based 
systems report the most outstanding features
in recognizing this attack, it still takes some human effort to figure out how
the three individual features can be stitched up to form an attack.

Regular expressions in 3) are used to describe patterns in each individual packet's
payload, which obviously can not handle session-level attacks. Besides, the widespread
use of encryption makes it harder to gain payload information. 
Type 4) provides a satisfying format. However, since state machine is not a very succinct
model, it has been able to work in very limited environments such as verifying 
protocols.

\subsection{The Case for NetQRE}

\sys{} synthesizes NetQRE~\cite{netqre} programs, which we will describe in more details in Section~\ref{sec:netqre}. We provide a high-level intuition here on why learning a NetQRE program from raw network traces can address all of the above challenges. NetQRE has an intuitive syntax for describing session-level patterns with quantities. For example, a Slowloris attack's pattern can be  programmed close to this natural language description:
A flow starts with a TCP handshake, followed by arbitrary packets whose intervals
are large. This pattern repeats many times.

At a high level, NetQRE is comprised of two parts:
{\em regular expression on packet sequence} and {\em quantitative aggregation}.  The regular expression part describes the properties of packets of interest
and their positional and repeating relations, similar to a plain regular expression
describing a string of characters.
Regular expressions are equivalent to finite state machines in expressiveness but
are typically more succinct, therefore they can easily capture the stateful nature of 
network protocols. Since NetQRE operates at the session-level, raw network traces 
can be directly used as inputs.

Based on the session-level regular patterns, the quantitative part of NetQRE
further specifies the following types of properties: how many times does a pattern
show up in the trace? Which sub-pattern shows up most frequently in the trace and
what is its frequency? If the trace is somehow to be split into sub-flows, how many
times will the pattern show up in each sub-flow? Such patterns may be concatenated
and nested based on the abstract syntax tree of the regular expression.
This gives NetQRE programs the additional ability 
to express quantitative patterns.

Similar to the probability score given by statistical models, the integer answer 
from the entire NetQRE program can be seen as how well the pattern is satisfied 
by the current input trace. Therefore, setting a learned threshold for this output answer
can turn the NetQRE program into a classifier, where network traces assigned
a higher value by the NetQRE program will be seen as a positive match. 

A more detailed comparison between our approach and other models is given in 
Table~\ref{table:model-comparison}

\begin{table}
 \footnotesize
 \begin{tabular}{m{6.5em} m{2.5em} m{2.5em} m{4em} m{4.8em} m{3em}} 
 \hline
 \textbf{Model} & Session-Level & Stateful & Explainable & Minimal Feature Engineering & Succinct \\ [0.5ex] 
 \hline
 \hline
 Raw Trace $\to$ \textbf{NetQRE} & \cmark & \cmark & \cmark & \cmark & \cmark \\ 
 \hline
 Raw Trace $\to$ \textbf{State Machine} & \cmark & \cmark & \cmark & \cmark & \xmark  \\
 \hline
 Feature Vector $\to$ \textbf{Heuristics} & \cmark & \xmark & \cmark & \xmark & \cmark \\
 \hline
 Feature Vector $\to$ \textbf{Statistics} & \cmark & \xmark & \xmark & \xmark & \xmark \\
 \hline
 Payload $\to$ \textbf{Regular Exp} & \xmark & \xmark & \cmark & \cmark & \cmark \\ [1ex] 
 \hline
\end{tabular}
\caption{Comparison between models}
\label{table:model-comparison}
\end{table}

\subsection{Search-based program synthesis}
\label{sec:background-synthesis}
Search-based syntax-guided synthesis is the process of finding a satisfying program based on
the grammar of the language and the specified logical constraints, which in our case
is the labelled input network traces. 

The main process of search-based synthesis is exploring the program space guided
by the grammar. We can use this simple regular expression grammar as an example:
\begin{lstlisting}
<re> ::= 0 | 1
		| <re> <re>
		| (<re>)*
\end{lstlisting}
We begin with the starting symbol or top-level non-terminal, which
is the $<re>$ symbol. In each step, we can expand any non-terminal in the program
following a rule in the grammar and get a new program. This process is called \textit{production}
or \textit{mutation}. For example, give the program $(<re>)*$, we have four
production choices over the $<re>$ symbol, which results in four new programs:$(0)*$
, $(1)*$, $(<re> <re>)*$ and $((<re>)*)*$. The first two results are complete and contain 
no non-terminals. Therefore they can be checked against the input examples. 
If either one works, we have found a solution and the search can stop. 
Otherwise we need to keep exploring the other two, 
which still contain non-terminals. 
We call these programs with non-terminals \textit{partial programs}.

One outstanding feature of search-based synthesis is that the only a priori knowledge
it needs is information about the language itself. No task-specific heuristics are required.

Although there has been a proliferation of research on program synthesis in recent years,
all the proposed techniques are specific to languages or type of languages. Popular synthesis
targets include string manipulation languages~\cite{gulwani2011automating,parisotto2016neuro,polozov2015flashmeta}, 
typed functional languages~\cite{osera2015type,polikarpova2016program},
simple imperative languages~\cite{so2017synthesizing}, languages that can be reduced to SMT solving problems~\cite{si2018learning}, etc.
NetQRE is a stream processing language with valued state. 
Efficiently synthesizing NetQRE program is the main technical challenge of our work.
No known technique can be directly applied to synthesis of NetQRE expressions.

In addition to that, fully describing properties of a malicious behavior, and especially
normal behaviors in contrast, requires learning from at least megabytes of network traffic, which is 
orders of magnitude larger than a typical program synthesis task addressed in existing
literature. Since one major overhead
of program synthesis is checking explored programs against the training
data, larger training data means proportional time consumption. For systems whose exploration
states grows exponentially to the input size~\cite{gulwani2011automating}, the situation could be even worse.
No known system has a design to especially address this concern. To overcome the difficulties, we design two synthesis techniques: partial execution and
merge search, which will be explained in detail in Section \S \ref{sec:synthesis}

\section{Overview}
\label{sec:overview}

\begin{figure*}[ht]
\centering
\includegraphics[width=0.8\textwidth]{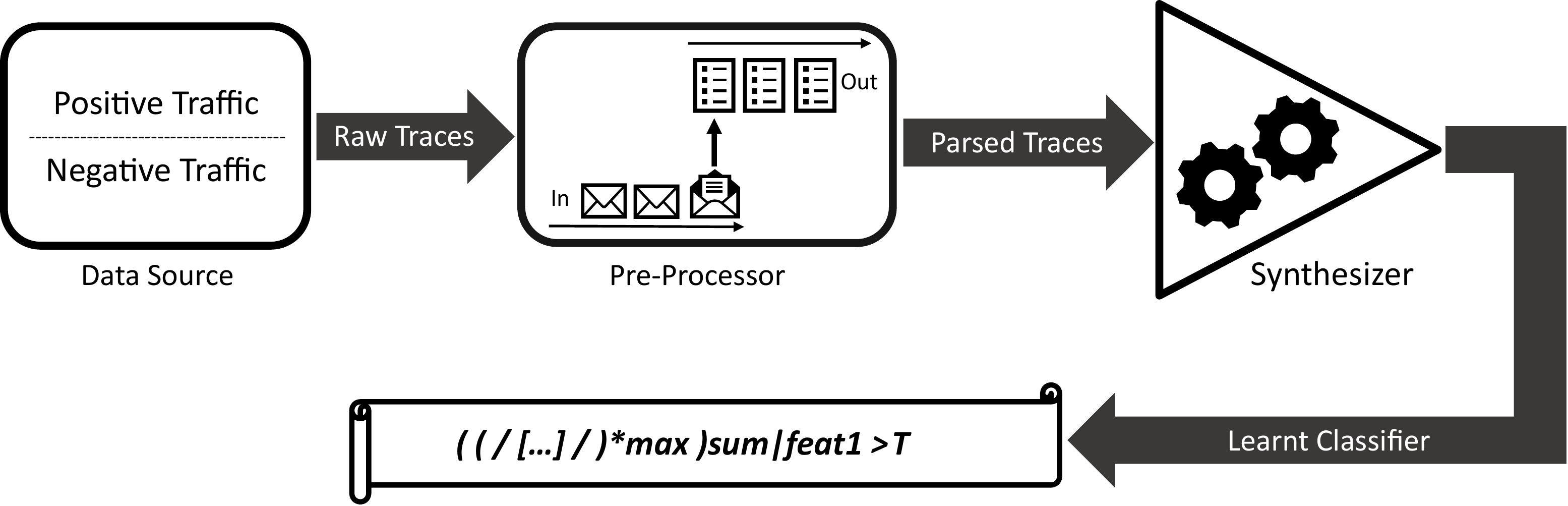}
\caption{System Overview}
\label{fig:overview-system}
\end{figure*}

Fig~\ref{fig:overview-system} shows the overall design of \sys{}.
\sys{}'s workflow is largely similar to a statistical supervised learning system, 
although the underlying mechanism is different. \sys{} takes labeled positive and 
negative network traces as input and outputs a classifier that can classify any 
new incoming trace.  To preserve the most of information from input data and
minimize the need of feature engineering, \sys{} considers three kinds of
properties in a network trace: (1) all available packet-level header fields, 
(2) position information of each packet within the sequence, (3) time information
associated with each packet.

Specifically, \sys{} represents a network trace as a stream of feature vectors: $S=v_0,v_1,v_2, \ldots$. 
Each vector represents a packet. Vectors are listed in timestamp order. 
Contents of the vector are parsed field values of that packet. 
For example, we can define 

$v[0] = \mathit{ip.src}$, $v[1]=\mathit{tcp.sport}$, $v[2]=\mathit{ip.dst}, \ldots$.

Depending on the information available, different sets of fields can be used to 
represent a packet. By default, we extract all header fields at the TCP/IP level. 
To make use of the timestamp information, we also append time interval since the previous
packet in the same flow to a packet's feature vector.
Feature selection is not necessary for \sys{}.


The output classifier is a NetQRE program $p$ that takes in a
stream of feature vectors. Instead of giving a probability score
that the data point is positive, it outputs an integer that quantifies
the matching of the stream and the pattern. The program includes a learnt threshold $T$. 
\sys{} aims to ensure that $p$'s outputs
for positive and negative traces fall into different sides of the threshold $T$.
Comparing $p$'s output for a data point with $T$ generates a label.
It is possible to translate $p$ and $T$ into executable rules, using a compilation step.

Given the above usage model, a network operator can use \sys{} 
to generate a NetQRE program trained to distinguish normal and suspected abnormal 
traffic generated from unsupervised learning systems. 
The synthesized program itself, as we will later show, forms the basis for 
deciphering each unknown trace. Consequently, traces whose 
patterns look suspicious can be subjected to a detailed manual analysis by 
the network operator. Moreover, the generated NetQRE programs can be further 
refined and compiled into filtering system's rules. The superior expressiveness
and explainability also allows easier maintenance and debugging in later tests
and deployment.



\section{Background on NetQRE}
\label{sec:netqre}

In this section, we give a more formal definition of NetQRE~\cite{netqre}. 
It is a high-level declarative language for querying 
network traffic. 
Streams of tokenized packets are matched against
regular expressions and aggregated by multiple
types of quantitative aggregators. The NetQRE language is defined by the following BNF grammar:
\begin{lstlisting}[basicstyle=\fontsize{8}{8}\ttfamily]
<classifier> ::= <program> > <value>

<program> ::= <split>

<split> ::= (<split>)<op>|<feats> 
            | <qre>
              
<qre> ::= (<qre> <qre>)<op>
          | (<qre>)*<op>
          | <unit>
          
<unit> ::= /<re>/

<re> ::= <re> <re>
         | (<re>)*
         | <pred>
         | _
              
<pred> ::= <pred> && <pred>
           | <pred> || <pred>
           | [<feat> == <value>]
           | [<feat> >= <value>]
           | [<feat> <= <value>]
           | [<feat> -> <prefix>]
                
<feats> ::= <feat>
            | <feats>, <feat>
<feat>  ::= 0 | 1 | 2 | ......
<op>    ::= max | min | sum
\end{lstlisting}

As an example, if we want to find out if any single source is sending more than 100 TCP packets,
the following expression describes the desired classifier:
\begin{lstlisting}[basicstyle=\fontsize{8}{8}\ttfamily]
( ( / [ip.type = TCP] / )*sum )max|ip.src_ip > 100
\end{lstlisting}

At the top level, there are two parts of it. A processing program that maps a network trace
to an output number:
\begin{lstlisting}[basicstyle=\fontsize{8}{8}\ttfamily]
( ( / [ip.type = TCP] / )*sum )max|ip.src_ip
\end{lstlisting}
and a threshold against which this value is compared:
\begin{lstlisting}[basicstyle=\fontsize{8}{8}\ttfamily]
> 100
\end{lstlisting}
They together form the classifier. Inputs fall into different classes based on the results of the comparison.

The processing program is designed to count the largest number of TCP packets any single
source is sending. The first step is to split the trace by source identifiers. 
This is achieved by \textit{flow split} ($<split>$):
\begin{lstlisting}[basicstyle=\fontsize{8}{8}\ttfamily]
( ............ )max|ip.src_ip
\end{lstlisting}
It splits the trace into sub-flows based on the value of the specified field (source IP address in this example). 
That is, packets sharing the same value in the field will be assigned to the same sub-flow. 
Then the matching and aggregation is done on each sub-flow individually. Finally, the expression
aggregates results from all sub-flows according to the aggregation operator (\textit{$<op>$})
(maximum in this example). In each sub-flow, we want to count the number of TCP packets. 
This can be broken down into three operations: 
(1) specifying a pattern that a single packet is a TCP packet, 
(2) specifying that this pattern repeats arbitrary number of times, and 
(3) adding $1$ to a counter each time this pattern is matched.

(1) is achieved by a \textit{plain regular expression} involving \textit{predicates}. A predicate 
describes properties of a packet that can match or mismatch one packet in the trace. 
Four types of properties frequently used in networks can be described:
\begin{enumerate}[topsep=5pt, itemsep=-0.3ex]
    \item \label{enum1:item1} It equals a value. For example: $[\mathit{tcp.syn} == 1]$
    \item \label{enum1:item2} It is no less than a value. For example: $[\mathit{ip.len} >= 200]$
    \item \label{enum1:item3} It is no greater a value. For example: $[\mathit{tcp.seq} <= 15]$
    \item \label{enum1:item4} It matches a prefix. For example: $[\mathit{ip.src\_ip} \rightarrow 192.168]$
\end{enumerate}
Predicates combined by concatenation and Kleene-star form a plain regular expression, 
which matches a network trace considered as  a string of packets. In this example,
we use a plain regular expression containing only one predicate that describes a single
packet whose protocol type equals TCP:
\begin{lstlisting}[basicstyle=\fontsize{8}{8}\ttfamily]
/ [ip.type = TCP] /
\end{lstlisting}
It is marked as a \textit{unit expression}, which indicates that this plain regular
expression should be viewed as atomic for quantitative aggregation. It either matches
a substring of the trace and outputs the value $1$, or does not match.

To achieve (2) and (3), we need a construct to both connect the plain regular expressions
to match the entire flow and also aggregate outputs bottom up from units at
the same time. We call it \textit{quantitative regular expression} ($<qre>$). 
In this example, we use iteration operator:
\begin{lstlisting}[basicstyle=\fontsize{8}{8}\ttfamily]
( / [ip.type = TCP] / )*sum
\end{lstlisting}
For matching, it works exactly like the Kleene-star operator describing that its sub-pattern
repeats arbitrarily many times. At the same time, for each repetition, the sub-expression's output
is aggregated by the aggregation operator. In this case, the sum is taken, which acts as
a counter for the number of TCP packets. The aggregation result for this expression will in turn
be returned as an output for higher-level aggregations.

The language also supports the concatenation operator:
\begin{lstlisting}[basicstyle=\fontsize{8}{8}\ttfamily]
(<qre> <qre>)<op>
\end{lstlisting}
which works exactly like concatenation for regular matching. It aggregates the quantity by applying
the $<op>$ on the outputs of two sub-expressions that match the prefix and suffix.


\smallskip
In addition to this core language, there is a specialization for synthesis purpose.
Enumerating all possible values for a field in the choice of predicates is computationally 
expensive for the synthesis algorithm. 
We observe that comparing a field with a value that does not appear in any of  the 
given examples will not
produce any meaningful information.
Therefore we use the relative position in the examples' value space instead of a specific value. 
For example, if the packets in the given examples have 4 different sequence numbers: 
$\{1, 3, 12, 15\}$, 
then $[\mathit{tcp.seq} >= 50\%]$ matches a packet with sequence number no less than 
50\% of values in the set. 
In other words, it is equivalent to $[\mathit{tcp.seq} >= 3]$. Prefix is also replaced by 
the common prefix of values in a range. For example, given the same value set above, 
$[\mathit{tcp.seq} \rightarrow [75\%, 100\%]]$ is equivalent to $[\mathit{tcp.seq} \rightarrow 11b]$ 
where $b$ means binary number ($12 = \textbf{11}00b, 15 = \textbf{11}11b$).
When the synthesis procedure finishes, the real values will be substituted into the
output program and it can run independently without the examples. 

\section{Synthesis Algorithm}
\label{sec:synthesis}

Given a set of positive and negative examples $E_p$ and $E_n$, 
respectively, the goal of our synthesis algorithm is to derive a NetQRE program $p_{f}$ 
and a threshold $T$ that differentiates $E_p$ apart from $E_n$. 
We start with notations to be used in this section:

\noindent{\bf Notation.} $p$ and $q$ denote individual programs, 
and $P$ and $Q$ denote sets of programs.  $p_1 \to p_2$ denotes it is possible to mutate $p_1$ 
following NetQRE's grammar to get $p_2$(see \S \ref{sec:background-synthesis} 
for definition of mutation). The relation $\to$ is transitive. 
There is a starting symbol $s_0$. All programs must be derived from  $s_0$, that is, $s_0 \to p$. 
We primarily care about the program part in the NetQRE classifier. 
Therefore $s_0$ is always a single non-terminal $<program>$.

$p(x)$ denotes program $p$'s output on input $x$, where $x$ is a sequence of
packets and $p(x)$ is a numerical value.
If $p$ is an incomplete program, i.e., if $p$ contains some non-terminal, 
then $p(x) = \{ q(x) \,|\, p \to q \}$ is a set of
numerical values, containing $x$'s output through all possible 
programs $p$ can mutate into. We define $p(x).max$ to be the maximum value in this set. 
Similarly, $p(x).min$ is the minimum value.

The synthesis goal can be formally defined as: 
$\forall e \in E_p, p_{f}(e) > T$ and $\forall e \in E_n, p_{f}(e) < T$. 


\subsection{Overview}
\begin{figure*}[ht]
\centering
\includegraphics[width=0.7\textwidth]{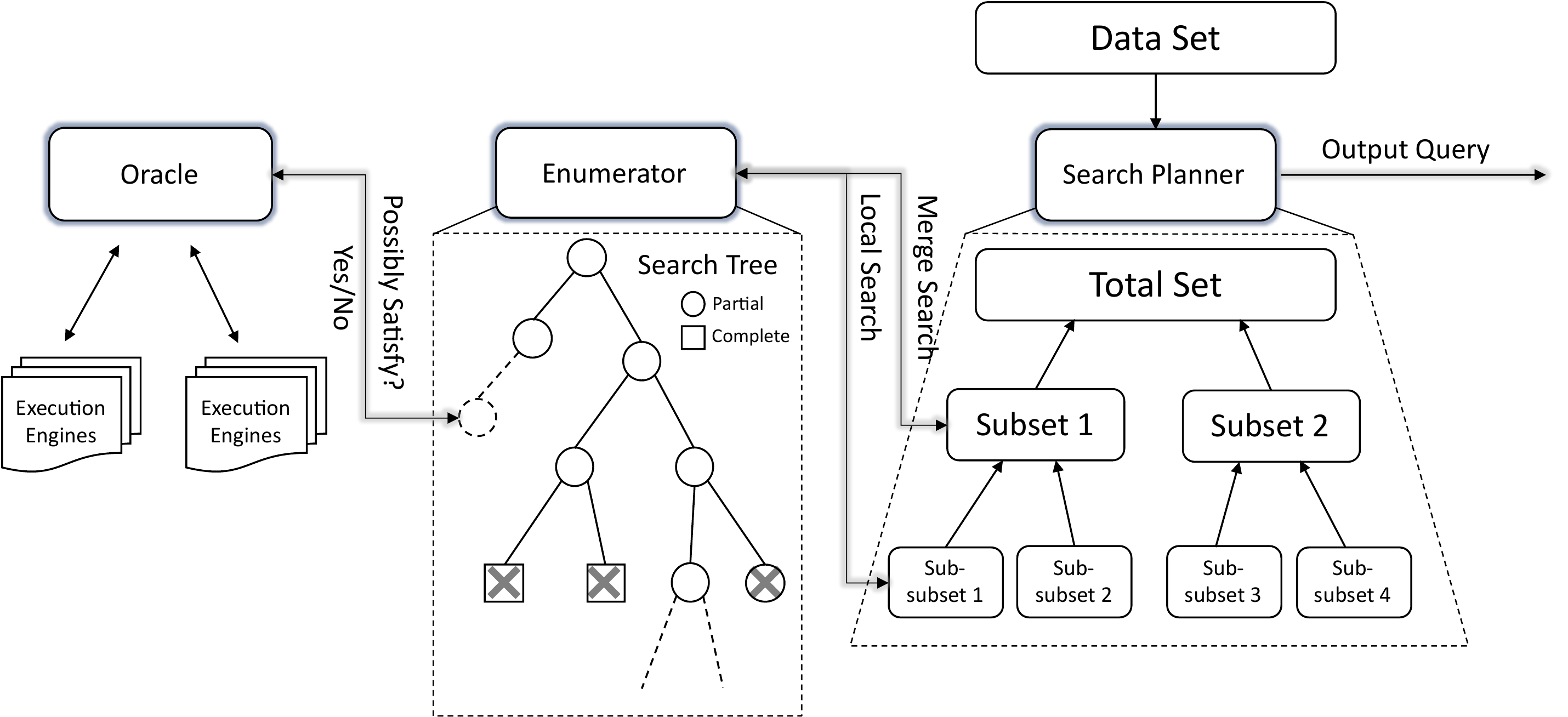}
\caption{Synthesizer Overview}
\label{fig:overview-synthesizer}
\end{figure*}

Our design needs to address two key synthesis challenges. First, NetQRE's rich grammar allows a large
possible program space and many possible thresholds for search. Second, the need to check
each possible program against a large data set collected from network monitoring 
tasks adds great overhead to the synthesis process.

We propose two techniques for addressing these challenges: {\em partial execution}
(Section~\ref{sec:partial}) and {\em merge search} (Section~\ref{sec:mergesearch}). 
Figure~\ref{fig:overview-synthesizer} shows an overview of the synthesizer. 

The top-level component is the \textit{search planner}, that assigns search tasks 
over subsets of the entire training data to the enumerator 
in a divide-and-conquer manner.
Each such task is a search-based synthesis instance, where the \textit{enumerator}  
enumerates all possible programs starting from $s_0$, 
expanded using the productions in NetQRE grammar, until one that can distinguish 
the assigned subset of $E_p$ and $E_n$ is found. 

The \textit{enumerator} optimizes for the first challenge by querying the distributed
\textit{oracle} about each partial program's feasibility and doing pruning early. 
The \textit{oracle} evaluates partial programs using {\em partial execution}.
The \textit{search planner} optimizes for the second challenge by merging search
results from subsets of the large training data, so as to save unnecessary checking, 
which we call the {\em merge search} strategy.

We next explain each technique in detail in the rest of this section.

\subsection{Partial Execution}
\label{sec:partial}

 A {\em partial program} is an incomplete program with non-terminals. 
In a standard syntax-guided synthesis process, the oracle only makes assertions
about complete programs. However, we observe that even if a program is partial, 
the existing skeleton already reveals some property about the set of possible completions. By making use of this information, accept/reject decisions can be made
earlier in the search process, thereby narrowing down the search to very few
valid branches in the search tree.

Specifically, we define $p(x) = \{\, q(x) \,|\, p\to q\,\}$ for a 
partial program $p$. It is obvious that the range of $p(x)$ contains the 
range of every $q(x)$. That is for all $q$ such that $p \to q$ , $p(x).min \le 
q(x).min \le q(x).max \le p(x).max$.
Making use of this knowledge, we may be able to decide if some completion $p_f$ such that
$p \to p_f$  can be a potential solution. If no such completion exists, we do not need to explore $p$ 
any further. On the other hand, we may be able to conclude that
 for all $q$ such that $p \to q$ , $q \in P_f$ holds, and in that case, 
we can select an arbitrary completion of $p$ to get a satisfying solution.
We will discuss how to make this decision later. For now, we describe 
an efficient way to evaluate $p(x)$ for a partial program $p$ and any input $x$.

\smallskip
\noindent{\bf Equivalent Completion}
A partial program $p$ with non-terminals cannot be directly evaluated on an input. 
We use an idea similar to overestimation in prior work~\cite{lee2016synthesizing,so2017synthesizing,so2018synthesizing}.
A completed version of $p$, denoted $\hat{p}$, is used to approximate $p$'s behavior. 
Recall that the necessary condition for early pruning is to have
for all $q$ such that $p \to q$ , $p(x)$ contains $q(x)$. 
If we can make sure that $\hat{p}(x)$ further contains $p(x)$,
then $p(x)$ can be approximated by $\hat{p}(x)$.

For many missing components in the syntax tree 
(that is, non-terminals that have not been expanded), it is straightforward to find
an equivalent completion. We replace (1) any uncertain numerical value with the 
largest or smallest possible value depending on the context, (2) any unknown predicate 
with $unknown$, (3) any unknown regular expression with $\_*$ and 
(4) any unknown quantitative regular expression with $(/\_\ \_*/)*sum$. 
The details of these four cases are briefly described below.

The syntax tree enumeration phase identifies proper numerical values by binary search. 
For example, to find $62.5\%$, the range $[50\%,100\%]$ is first explored. 
If it works, this is refined to $[50\%, 75\%]$, and eventually $62.5\%$. 
If there is an incomplete predicate $[\mathit{feat1} >= [50\%,100\%]]$, 
it can be completed by taking the smallest possible value $50\%$ in the
unknown part and turned into $\hat{p}$: $[\mathit{feat1} >= 50\%]$. 
If the operator is $<=$, we take the largest possible value instead.

We define $unknown$ as 
a Boolean state that is possibly true and possibly false, which indicates the
uncertain status of a partial predicate with incomplete parts other than numerical values.
Wherever a $true$ is required, $unknown$ also works, since it already 
implies the possibility of matching. The calculation rule for $unknown$ is:
\begin{lstlisting}
    T = true, F = false, U = unknown
    T && U = U  T || U = T
    F && U = F  F || U = U
    U && U = U  U || U = U
\end{lstlisting}

The remaining two cases are both within the grammar of NetQRE.
$\_*$ matches an arbitrary number of arbitrary packets, therefore 
containing the matching results of all possible regular expressions.
$(/\_\ \_*/)*sum$ matches an arbitrary
number of packets and outputs 0 (in case no packet is matched) 
or $[1,n]$ where $n$ is the number of packets matched. 
Since a unit expression outputs a constant 1 and
there is no multiplicative aggregator, this is exactly the range of all
possible outputs of all possible expansions.

There are some non-terminals that cannot be completed in this way, such as
flow split and aggregation operators. We put a complexity
penalty over these non-terminals if they are not expanded, therefore
encouraging expanding them earlier to allow partial execution.
Since flow split is not frequently used, and aggregation operator
does not have many expansion choices, they are not a major source 
of overhead.

\smallskip
\noindent{\bf Computing Ambiguity:}
Notice that although the finished program still largely follows NetQRE's
grammar, its semantics is different, because this process introduces
ambiguity into the program. That is, for a given input $x$, the completed
program $\hat{p}$ can have different matching strategies and different
outputs. In core NetQRE, such a case would have produced a \textit{conflict} 
output. But in partial execution, our goal, and also the main challenge, 
is to properly estimate the set of all possible outputs for the possible 
completions.

We demonstrate this ambiguity problem by an example. Suppose there are two predicates
$A$ and $B$ defined as:
\begin{lstlisting}
A: [ip.type == TCP]
B: [ip.type == UDP]
\end{lstlisting}
We can write a NetQRE program based on them:
\begin{lstlisting}
( ( /AA/ )*sum ( /B/ )*sum )max
\end{lstlisting}
which describes a trace with an even number of TCP packets followed by an arbitrary
number of UDP packets. It counts the number of TCP packet pairs and the number of 
UDP packets, and outputs the larger number. Since $A$ and $B$ are mutually exclusive,
the expression is not ambiguous. 

\begin{figure}[ht]
\centering
\includegraphics[width=0.5\textwidth]{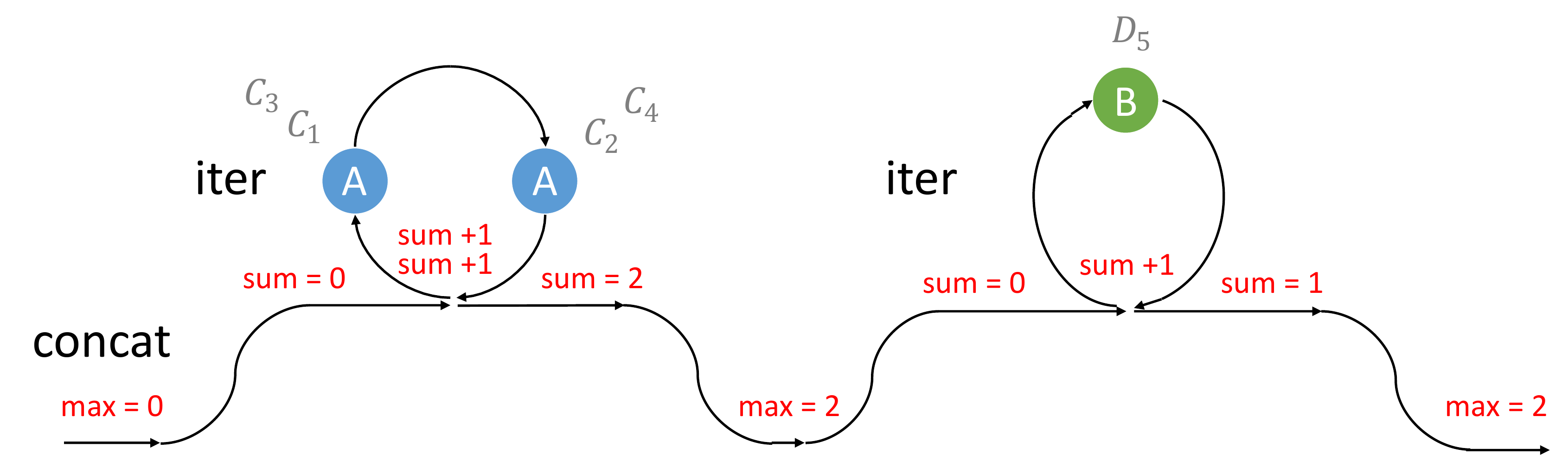}
\caption{Illustration of an unambiguous program}
\label{fig:flowchart1}
\end{figure}

Suppose there is a trace of packets $CCCCD$, where $C$'s are some TCP packets and 
$D$ is a UDP packet. Let us first consider how it is processed by the unambiguous 
NetQRE program. The execution can be illustrated by the flowchart in Figure~\ref{fig:flowchart1}.
A trace first goes to the left cycle that consumes pairs of TCP packets, and then 
to the right for UDP packets. The order of packets matched is also shown in the
figure as subscripts. 

In order to compute the numerical output, the program needs to maintain aggregation 
states during the matching. For example, when it processes the third
packet $C_3$, the execution is at the middle of the second loop of the left cycle. 
The accumulated sum of this iteration cycle is currently $1$, and maximum value of 
the outermost concatenation is currently the initial value $0$. 
These are core states for computing ambiguous outputs.
Eventually, the outermost aggregation result $2$ is taken as the final output.

\smallskip
Now let us look at the synthesis steps. Suppose that during the search, 
we have explored part of this program and the predicate $B$
is not yet known:
\begin{lstlisting}
( ( /AA/ )*sum ( /<pred>/ )*sum )max
\end{lstlisting}
To evaluate this partial program, we complete it by replacing
the missing predicate with $unknown$, denoted as $\_$ below, which matches any packet:
\begin{lstlisting}
( ( /AA/ )*sum ( /_/ )*sum )max
\end{lstlisting}

As a result, the program has become ambiguous. To evaluate it on the input trace $CCCCD$,
there are three different correct matching strategies: matching the first
iteration of two TCP packets with 0, 2, or 4 number of $C$ packets respectively, and matching
the rest of the trace with the iteration of wildcard. They produce three outputs: 5, 3, 2.
The set $\{2, 3, 5\}$ is an optimal result. But in practice, since we 
will compare with a specific threshold, only the upper-bound and lower-bound 
of all possible outputs is needed. Therefore we want the output to be the interval
$[2,5]$.

A strawman method is simply to enumerate all possible matching strategies and
take the union of all their outputs. The problem is that there can be exponentially
many distinct matching strategies with respect to the length of the network trace, 
leading to unacceptable synthesis time.

We solve this problem by an approximation: we merge "close" matching strategies.
Two strategies are defined to be "close" if at some step of their matching process
(1) they have matched the same number of packets in the trace and (2) the last
predicate they have matched is exactly the same. We explore all matching strategies
simultaneously and do a merging whenever two strategies can be identified to be
close.

\begin{figure}[ht]
\centering
\includegraphics[width=0.5\textwidth]{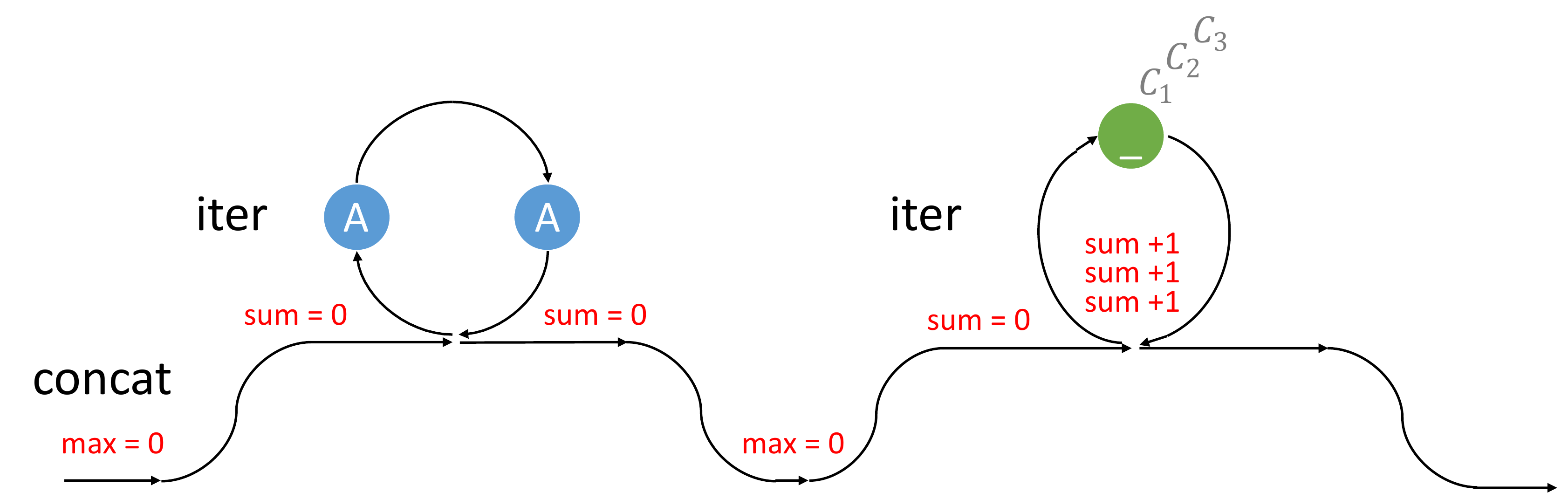}
\caption{Illustration of the first 3 steps of strategy one}
\label{fig:flowchart2}
\end{figure}

\begin{figure}[ht]
\centering
\includegraphics[width=0.5\textwidth]{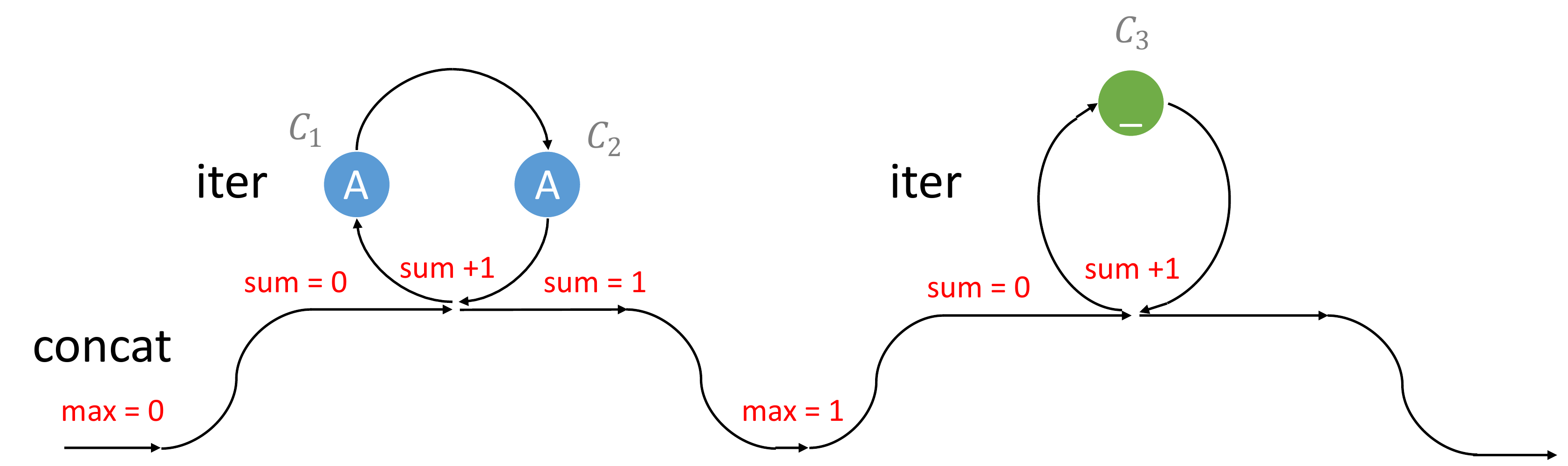}
\caption{Illustration of the first 3 steps of strategy two}
\label{fig:flowchart3}
\end{figure}

We now describe how this merging works. Again, we use the program and trace above
as an example. We inspect the two matching strategies that match the first iteration 
of two TCP packets against 0 and 2 packets of the trace respectively. After matching the third
packet $C$ against the wildcard $\_$, they can be identified to be close. 
Their matching and aggregation states before this point are shown in Figures~\ref{fig:flowchart2} and \ref{fig:flowchart3}.

\begin{figure}[ht]
\centering
\includegraphics[width=0.5\textwidth]{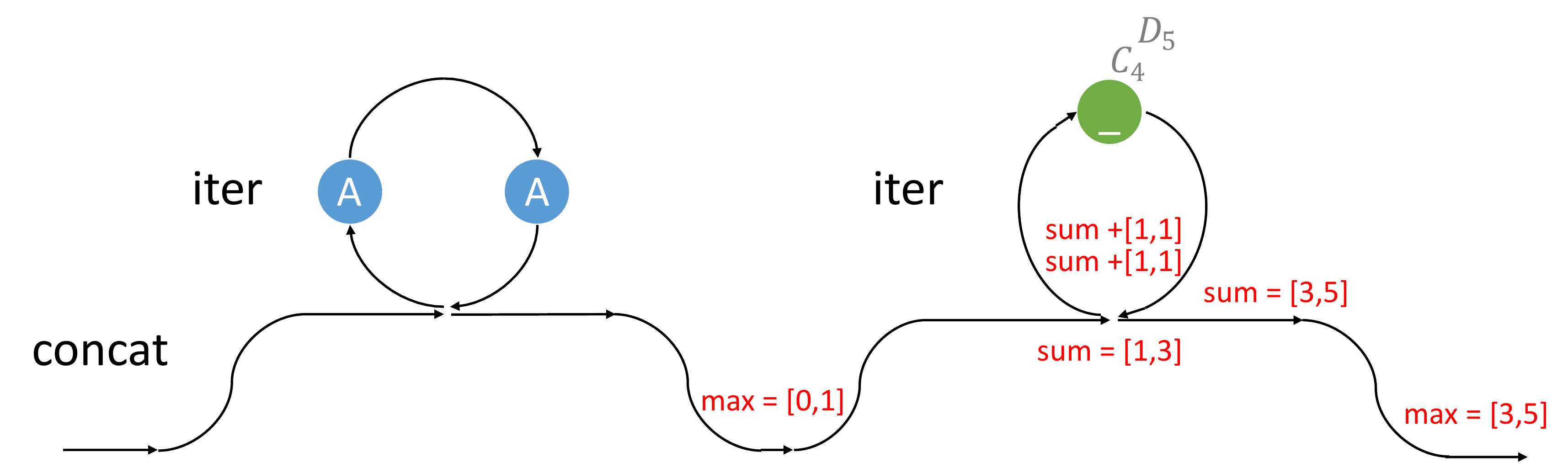}
\caption{Illustration of the last 2 steps of merged strategy one \& two}
\label{fig:flowchart4}
\end{figure}

We observe that the corresponding executions have maintained different aggregation states. For strategy
one, the current maximum for outermost concatenation is $0$ and the current sum for the second iteration is 3.
For strategy two, the current maximum for the concatenation is $1$ while the current sum for the
second iteration is 1. Similar to the way we handle final outputs, we merge these aggregation states
by recording the range of all possible values from merged strategies. In this specific
example, the two values will result in intervals $[0,1]$ and $[1,3]$ respectively.
For the remaining part of the matching, the aggregations will be done on the intervals,
which is illustrated in Figure~\ref{fig:flowchart4}. 
Eventually, the final aggregation result $[3,5]$ is the estimated output
for this merged strategy.

The regular matching result by this approximation is correct,
since at any step, a matching strategy's remaining part is determined 
by matched packets' length and the current predicate.
It can also be proven by the properties of interval arithmetic 
that the aggregation result strictly contains the true output range. 
Or more formally, $\hat{p}(x).min \le p(x).min \le p(x).max \le \hat{p}(x).max$.
Therefore $p(x)$ can be approximated by $\hat{p}(x)$.

Intuitively, the proposed evaluation scheme works well because we only care about the boundary of outputs, which
are represented by intervals as the abstract data type. We implement the execution and 
approximation process by the Data Transducer model proposed by \cite{alur2019modular},
which consumes a small constant memory and liner time to the input trace's length 
given a specific program.

\smallskip
\noindent{\bf Make Decision:} Given that we are able to evaluate $\hat{p}(x).min$ 
and $\hat{p}(x).max$, the next step requires us to make the accept/drop decision 
and find the proper threshold $T$.

Let $q$ be a complete program and assume there is only one pair of examples
$e_p$ and $e_n$. Now that we can evaluate ambiguous programs, we allow
$q$ to be ambiguous too. For $q$ to accept $e_p$ and $e_n$, there must
be a threshold $T$ such that $q(e_n).max < T < q(e_p).min$. Therefore,
given a pair of examples $e_p$ and $e_n$, a program $q$ is correct if and
only if $q(e_n).max < q(e_p).min$. When this holds, any value between
$q(e_n).max$ and $q(e_p).min$ can be used as the threshold.

\smallskip
\noindent
\textbf{Lemma 1}: There exists a correct program $q$ such that $p \to q$
only if $\hat{p}(e_n).min < \hat{p}(e_p).max$

\noindent
\textbf{Lemma 2}: If $\hat{p}(e_n).max < \hat{p}(e_p).min$ then any
program $q$ such that $p \to q$ is correct.

\begin{figure}[ht]
\centering
\includegraphics[width=0.4\textwidth]{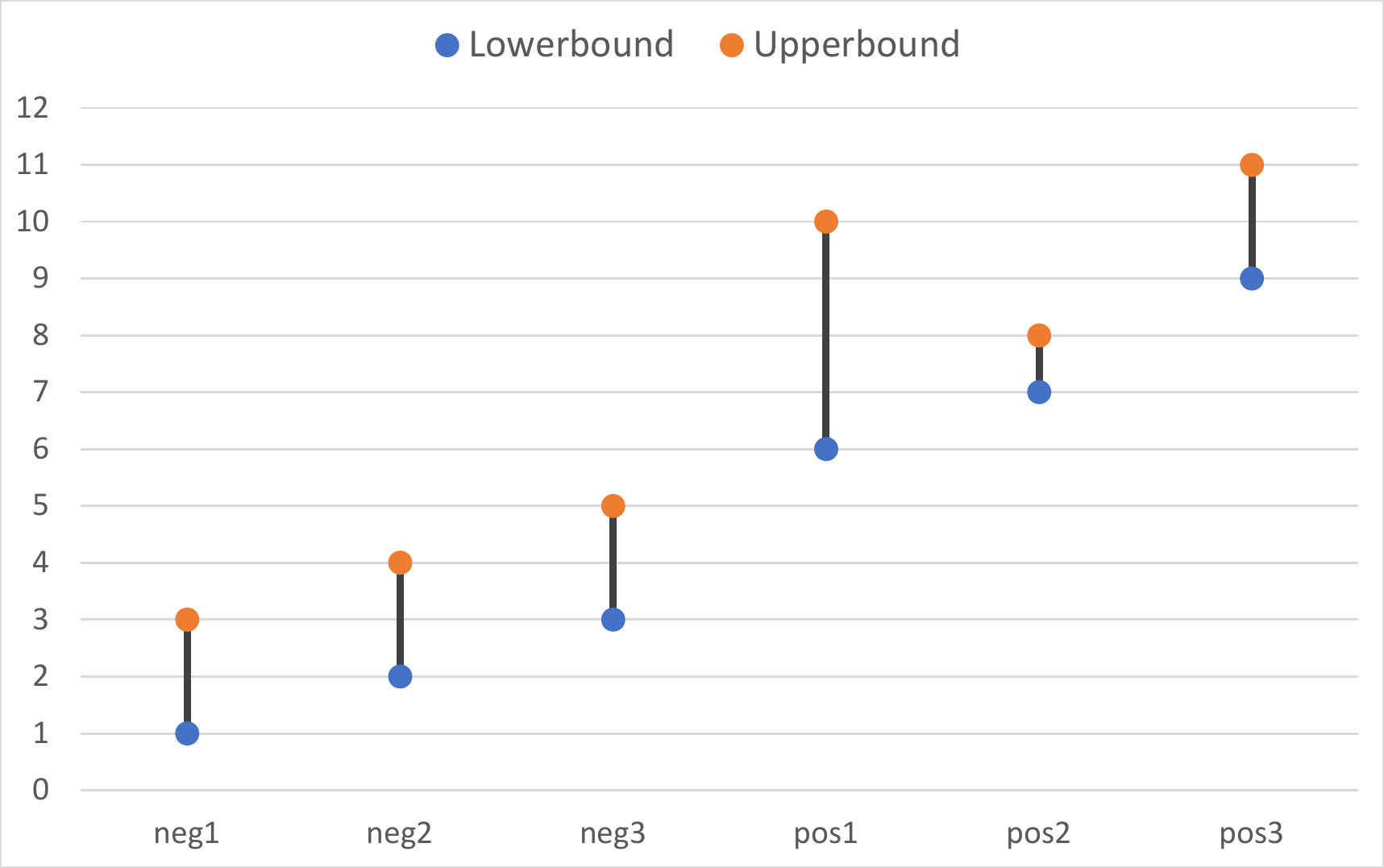}
\caption{A correct program found. No negative
output can ever be greater than any positive output.
5.5 can be used as a threshold}
\label{fig:threshold-good}
\end{figure}

\begin{figure}[ht]
\centering
\includegraphics[width=0.4\textwidth]{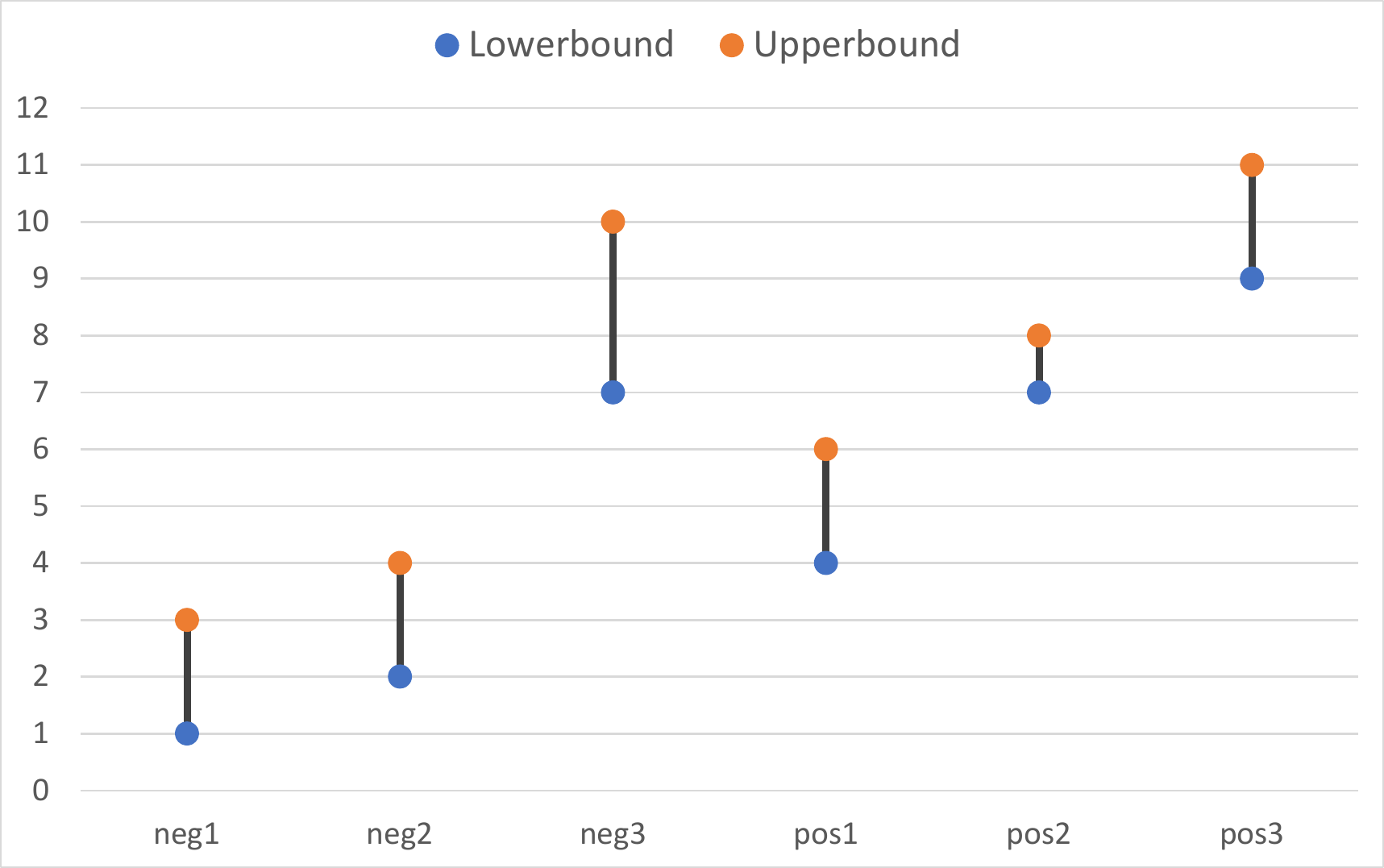}
\caption{A bad program. pos 1 can never be
greater than neg 3.}
\label{fig:threshold-bad}
\end{figure}

\smallskip\noindent
From Lemma 1, we can decide if $p$ must be rejected. From Lemma 2, we can decide if $p$
must be accepted. These criteria can be extended to more than 1 pair of examples. 
Figures~\ref{fig:threshold-good} and \ref{fig:threshold-bad}
show two intuitive examples for explanations
of the decision making process (but do not necessarily represent properties of real data sets). 
Each vertical bar represents the output
range of the corresponding data point produced by the program under investigation. 

\subsection{Merge Search}
\label{sec:mergesearch}


In the rest of this subsection, we describe three heuristics for scaling up synthesis to large data sets, 
namely {\em divide and conquer}, {\em simulated annealing}, and {\em parallel processing.}
We call the combination of these the 
 {\em merge search} technique.

\noindent{\bf Divide and Conquer.} We process large data sets in a divide-and-conquer manner. Enumerating and verifying
programs on large data sets is expensive. If we can learn patterns on small subsets
and merge them into a global pattern with low overhead, then the performance will be 
significantly improved. This strategy works well for many pattern learning tasks due to two 
observations.

First, the pattern of the entire data set is usually shaped by a few extreme data points.
Most data points are highly similar and provide almost no additional information.
For example, if we are learning the SYN flood attack, one complete TCP handshake is enough to tell that 
a valid TCP connection does not belong to SYN flood. All other complete TCP handshakes 
provide no more information. Therefore, looking at these 
extreme data points locally is enough to figure out critical
properties of the global pattern. A larger data set has a higher probability to 
cover more extreme data points, and thus raising the accuracy. 
But learning by the entire data set is unnecessary.

Second, resolving conflicting patterns is straightforward. Only in rare cases will
they fundamentally conflict with each other. Mostly, the local patterns are simply describing 
different aspects of the same object, e.g. "a broken handshake" and "a
large number of flows" in the SYN flood example. Simple merge operations such as 
disjunction, truncation or concatenation are enough to unify them.

This divide and conquer strategy is captured in the following algorithm:
\begin{lstlisting}
def d&c(dataset)
    if dataset.size > threshold
        subsetL,subsetR = split(dataset)
        candidateL = d&c(subsetL)
        candidateR = d&c(subsetR)
        return merge(dataset, candidateL, candidateR)
    else
        return synthesize(dataset, s0)
\end{lstlisting}

The "split" step corresponds to evenly splitting positive and negative examples.
The conquer, or "merge" step is more complex.
First, each candidate expression is turned into its abstract syntax tree (AST). Then, low-level subtrees 
of these ASTs are collected. That is, suppose there is a low-level non-terminal
$a$ in the AST, and the sub-program corresponding to $a$ is $b_1$ $b_2$ $b_3$... $b_m$
where each $b_i$ is a terminal. We then remove the entire subtree of $a$ from the
AST and add the rule $<a> ::= b_1 b_2 b_3...b_m$ to the syntax. This creates 
a shortcut for search in the future. 

Using the SYN flood attack as an example. Suppose we have found a local pattern:
\begin{lstlisting}
    ( /_ [tcp.syn=1] && [tcp.ack=1]/ )sum|flow_id
\end{lstlisting}
and would like to benefit from the discovery of the complex predicate in the middle. This 
is achieved by adding a new rule to NetQRE's syntax:
\begin{lstlisting}
<pred> ::= ......
                |......
                |[tcp.syn=1] && [tcp.ack=1]
\end{lstlisting}
We are then left with a partial program:
\begin{lstlisting}
    ( /_ <pred>/ )sum|flow_id
\end{lstlisting}

This process can be recursively done
on all non-terminals in $a$'s subtree. The partial programs left over are used 
as seeds for synthesizing future programs, and a new search on the merged data set will be done starting with these seed programs.
With the expanded syntax, all previously learnt local patterns can now
be used as building blocks for the global pattern. To encourage reuse of
local patterns, the seeds and added syntax nodes are given complexity rewards
to raise their priority. 

The pseudocode for merge is also given below:
\begin{lstlisting}
def merge(dataset, candidateL, candidateR)
    candidates = [candidateL, candidateR]
    for p in candidates
        if p.accept(dataset)
            return p
    for p in candidates
        harvest(p)
    return synthesize(dataset, candidates)
    
def harvest(ast)
    if ast.depth <= depth_threshold
        ast.root.syntax().add_option(ast)
    for t in ast.subtrees
        harvest(t)
    if ast.depth == depth_threshold
        ast.subtrees.clear()
\end{lstlisting}

In practice, many search results can be directly reused from cached results generated from previous
tasks on similar subsets. This optimization can further reduce the synthesis time.

\noindent{\bf Simulated Annealing}
When searching for local patterns at lower levels, we require the Enumerator 
to find not 1 but $t$ candidate patterns for each subset. Such searches are fast for 
smaller data sets and can cover a wider range of possible patterns. 
As the search 
goes to higher levels for larger data sets, we discard the least accurate local 
patterns and also reduce $t$. The search will focus on refining the currently optimal
global pattern. This idea is based on traditional simulated annealing algorithms and helps to improve the synthesizer's performance in many cases.

\noindent{\bf Parallelization.} Most steps in the synthesis process are inherently parallelizable. They include \cir{1} doing synthesis
on different subsets of data, \cir{2} exploring different programs in the enumeration, 
\cir{3} verifying different programs found so far, \cir{4} executing a program on different data points during the verification.

We focus less on optimizing \cir{1} and \cir{2} since they are not the performance bottlenecks. We instead focus on parallelizing \cir{3} and \cir{4} over multiple cores. In our implementation,  using 5 machines with 32 cores
each, we devote one thread each to run task \cir{1} and \cir{2} on one machine, 64 threads on the same machine to run
task \cir{3}, and 512 threads distributed over the remaining four machines to run task \cir{4}.
The distributed version is approximately two orders of magnitude faster
than the single-threaded version for complex tasks. Given more computing power, a proportional speedup
can be expected.

\section{Evaluation}
\label{sec:eval}

We implemented \sys{} in 10K lines of C++ code. Our experiments are carried out in a cluster
of five machines directly connected by Ethernet cable, each with 32 Intel(R) Xeon(R) E5-2450 CPUs.
The frequency for each core is 2.10GHz. The core \sys{} synthesizer runs on one machine,
with 64 threads exploring new programs and doing early syntactical checks.
512 execution engine threads are distributed over the remaining four machines.
Execution tasks are assigned using RPC. Our evaluation summary is: 

\begin{enumerate}[topsep=5pt, itemsep=-0.2ex]

\item \sys{} requires minimal feature engineering and data preparation work (Section~\ref{sec:prep}); 

\item \sys{} generates NetQRE programs that achieve comparable accuracy as state-of-the-art network traffic classification systems (Section~\ref{sec:eval-learning}); 

\item \sys{} generates NetQRE programs that are natural to decipher, edit and tune by a network operator(Section~\ref{sec:tuning});

\item Generated NetQRE programs are amenable to efficient execution and deployment on existing monitoring systems, for example, compilation into Bro~\cite{paxson1999bro} programs (Section~\ref{sec:deploy}); 

\item Synthesis algorithms in \sys{} are efficient and can be practically deployed. In several instances, the optimizations ensure that synthesis time is a few minutes, and complete
classification tasks that are otherwise impossible to finish within reasonable time limits (Section~\ref{sec:synthesistime}).

\end{enumerate}

\subsection{Data Preparation}
\label{sec:prep}

We utilize the CICIDS2017 database\cite{sharafaldin2018toward}, a public repository of attack traffic 
used for evaluating intrusion detection systems. The database contains five days worth of 
network traffic consisting of benign and a wide range of 
attack traffic. The data set consists of labeled data in the conventional format of extracted feature
vectors of each flow. Each flow is given a label of "benign" or a specific attack type. 

Since NetQRE operates on raw traffic, for our experiments, 
we pre-process the data to correlate the feature vectors' information and labels with the raw 
traffic data, and use the labeled raw packet flows for our experiments.

In our experiments, we utilize training and testing data
for eight types of attacks. These attacks are based on botnets, Denial of service (DoS), 
port scanning, and password cracking. We learn each type of attack against benign traffic separately. 
To use as many data as possible, for each attack type, we use 1500 positive (attack) flows 
and 10000 negative (benign) flows for training, and another distinct data set of similar 
size for testing. Note that "training" in our case refers to generating the NetQRE expression 
from examples. 

The main benefit of \sys{} in this experiment is the {\em minimal need} for feature engineering. We simply use all header fields of TCP and IP, and as additional features, we add the inter-packet arrival time between adjacent packets in the same flow. In total, there are 19 features per packet. Given a trace of $N$ packets, the feature matrix presented to \sys{} is of size $N \times 19$. In most of our experiments, this consists of thousands of features. 

\sys{} data preparation work requires minimal feature engineering. In contrast, other state-of-the-art systems rely on 
a carefully designed feature extraction step to work well. For example, the feature vectors
included in CICIDS2017 database contain 84 features extracted by the CICFlowMeter\cite{draper2016characterization,lashkari2017characterization} tool for each flow, characterizing performance metrics of the entire flow such as duration, mean forward packet length, min activation time, etc. Kitsune\cite{mirsky2018kitsune} extracts bandwidth information
over the past short periods as packet-level features. DECANTeR\cite{bortolameotti2017decanter} uses HTTP-level properties such as constant header fields, language, amount of outgoing information, etc. as flow-level features.  
\sys{} avoids complex feature engineering, and this reduces errors and data preparation time. 
As we will later demonstrate, despite minimal feature engineering, \sys{} is able to achieve comparable accuracy as competing approaches.

\subsection{Learning Accuracy}
\label{sec:accurate}
\label{sec:eval-learning}

We next validate \sys{}'s learning accuracy using the following evaluation methodology:

\begin{itemize}
\item For each attack from the CICIDS2017 data set, we use the training data (attack and normal traffic) 
as input to \sys{} to learn a NetQRE program. The full set of synthesized NetQRE programs is shown in Appendix~\ref{appendix:A}. The NetQRE program is then validated on the testing set for accuracy. 
Note that we focus on single attacks in our experiments. Generating NetQRE programs from 
traces consisting of multiple attacks is a direction of future research.

\item The output of \sys{} includes a NetQRE program that maps a network trace to an integer output
and a recommended range for the threshold. By modifying the threshold, true positive rate (TP) and false positive
rate (FP) can be adjusted, as we will later explain in Section~\ref{sec:post-processing}.
We use  AUC (Area under Curve) - ROC (Receiver Operating Characteristics) metric, 
which is a standard statistical measure of classification performance. 

\item In our evaluation, \sys{} computes the top five candidate programs instead of one, 
and the program with the highest test accuracy is picked as the final answer. 
In practice, all the top candidates are presented to the network operator to choose, 
based on domain knowledge. 
\end{itemize}

\begin{figure*}[ht]
\centering
\includegraphics{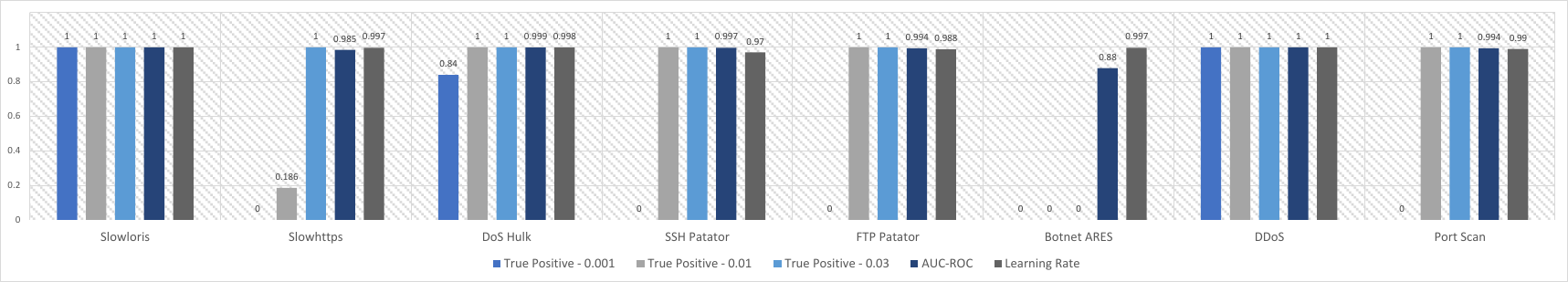}
\caption{\sys{}'s true positive rate under low false positive rate, AUC-ROC and learning rate for 8 attacks in CICIDS2017 (higher is better)}
\label{fig:aucroc}
\end{figure*}

Figure~\ref{fig:aucroc} contains results for eight types of attacks. Apart from AUC-ROC values, 
we also show the true positive rates when false positive rate is adjusted to 3 different levels:
$0.001$, $0.01$, and $0.03$. Given that noise is common in most network traffic, 
the last metric shown in Figure~\ref{fig:aucroc} is the highest achievable learning rate, 
which is defined as the ratio of training examples the learnt classifier can
correctly classify.

Overall, we observe that \sys{} performs well across a range of attacks
with accuracy numbers on par with prior state-of-the-art systems such as Kitsune,
which has an average AUC-ROC value of $0.924$ on nine types of IoT-based attacks,
and DECANTeR, which has an average detection rate of $97.7\%$ and a
false positive rate of $0.9\%$ on HTTP-based malware.
In six out of eight attacks, \sys{} achieves above $0.994$ of AUC-ROC and $100\%$ of true 
positive rate at $1\%$ false positive rate. The major exception is Botnet ARES, 
which consists of a mix of malicious attack vectors. 
Handling such multi-vector attacks is an avenue for our future work.

\subsection{Post-processing and Interpretation}
\label{sec:post-processing}
\label{sec:tuning}

One of the benefits of \sys{} is that it generates an actual classification program that 
can be further adapted and tuned by a network operator. The program itself is also 
close to the stateful nature of session-layer protocols and attacks, and thus is
readable and provides a basis for the operator to understand the attack cause. 
We briefly illustrate these capabilities in this section.

\begin{figure*}[!htb]
\minipage{0.32\textwidth}
\includegraphics[width=\linewidth]{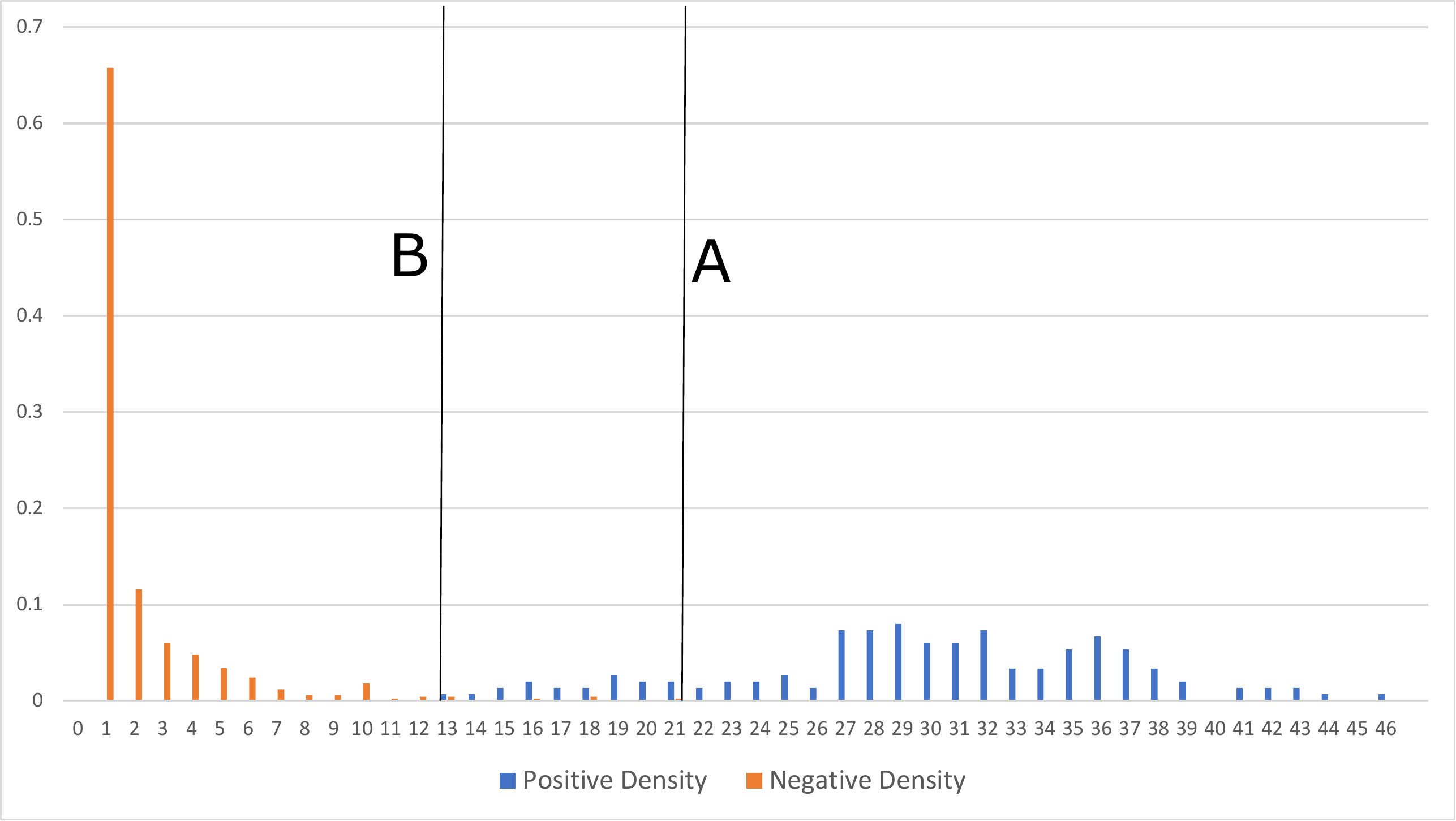}
\caption{Output distribution of training set(DoS Hulk)}
\label{fig:output-train}
\endminipage\hfill
\minipage{0.32\textwidth}
\includegraphics[width=\linewidth]{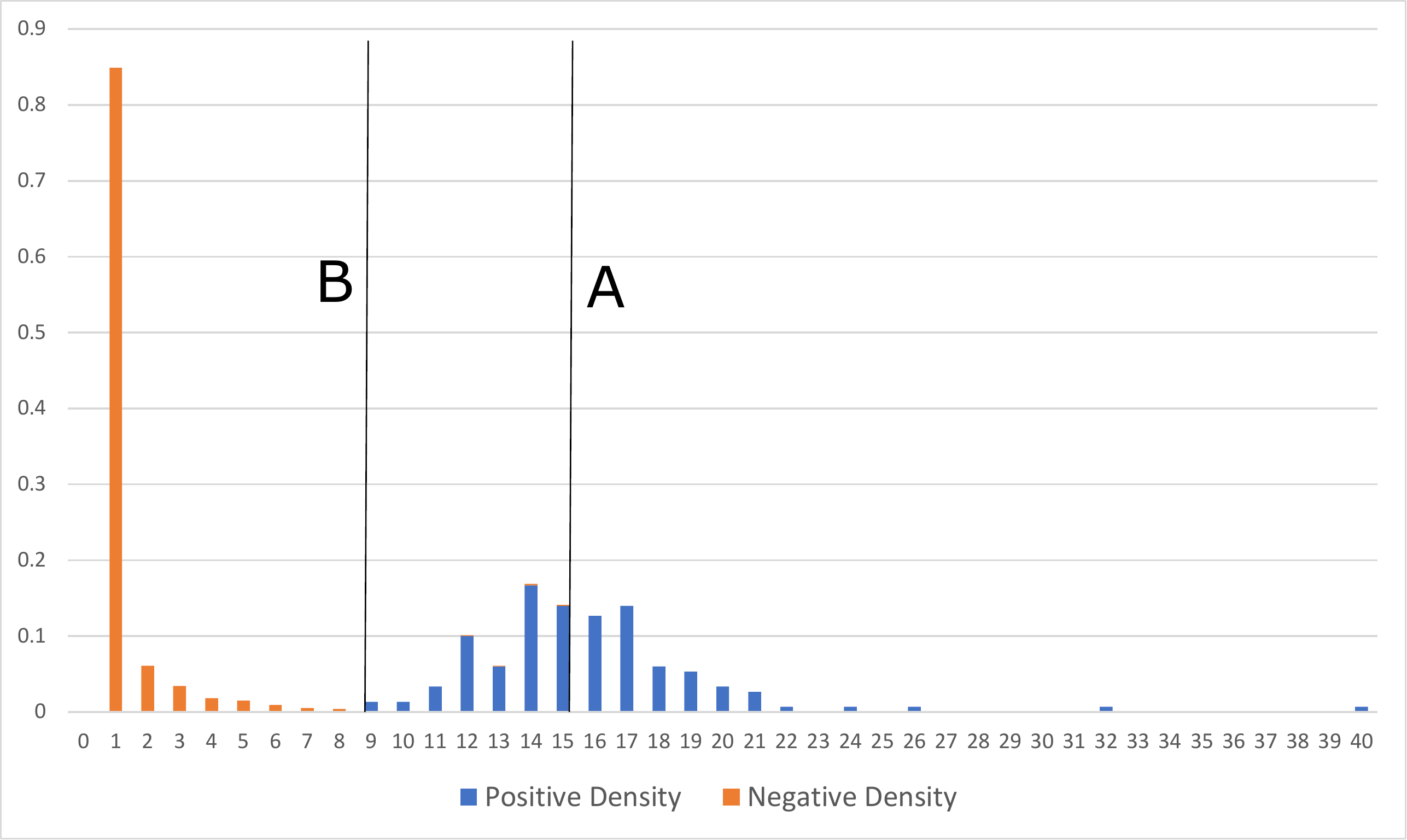}
\caption{Output distribution of test set(DoS Hulk)}
\label{fig:output-test}
\endminipage\hfill
\minipage{0.32\textwidth}
\includegraphics[width=\linewidth]{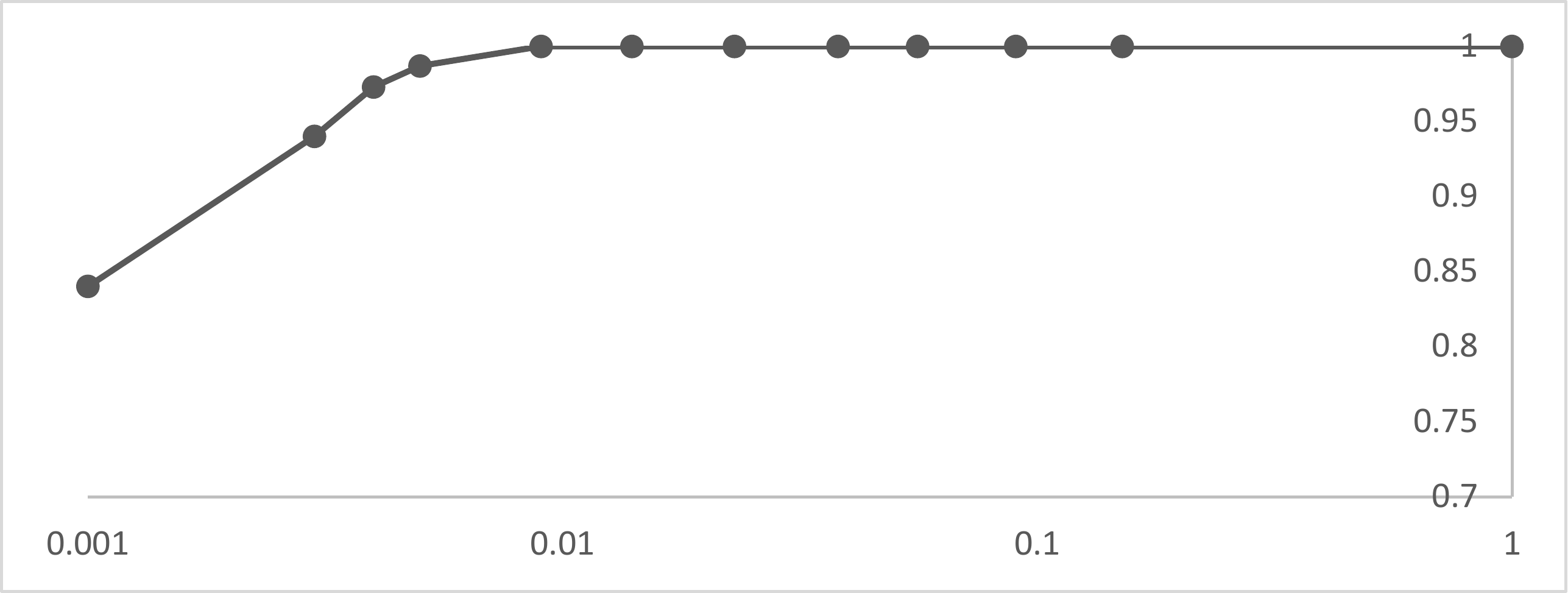}
\caption{ROC Curve, logarithmic scale(DoS Hulk)}
\label{fig:roc}
\endminipage
\end{figure*}

\noindent \textbf{FP-TP Tradeoff}
Network operators need to occasionally tune a classifier's sensitivity to false positives
and true positives. \sys{} generates a NetQRE program with a threshold $T$. This threshold can be adjusted to vary the false positive and 
true positive rate. 
Figures~\ref{fig:output-train} and \ref{fig:output-test} show 
the output distribution from positive and negative examples in the DoS Hulk attack. 
$A$ denotes the largest negative output and $B$ denotes the smallest positive output.
When $A > B$, there is some unavoidable error. If the training data has such an error, the data points between $A$ and $B$
are treated as noise. We can slide the threshold $T$ from $B$ to $A$  and obtain an ROC curve for the test data, as illustrated in Figure~\ref{fig:roc}.

\noindent \textbf{Interpretation}
We describe a few learnt NetQRE programs to demonstrate how a network operator can interpret the classifiers. A full list of our classification programs is shown in Appendix~\ref{appendix:A}. The NetQRE program synthesized by \sys{} for DDoS is:

\begin{lstlisting}
( ( /_* A _* B _*/ )*sum /_* C _*/ )sum > 4
Where 
A = [ip.src_ip->[0%,50%]]
B = [tcp.rst==1]
C = [time_since_last_pkt<=50%]
\end{lstlisting}

The DDoS attacker launches a SYN or ACK flood attack from a botnet of machines to exhaust memory resources on the victim server.
The above program is generated from the actual DDoS attack traffic. The detected pattern consists of packets that start with 
source IP in the lower half of the value space, followed by a packet with the reset bit set to 1, with arbitrary packets in-between.
This pattern is then followed by a packet with a short time interval from its predecessor. Finally, the program considers 
the flow a match if the patterns show up with a total count of over $4$. 

The synthesized pattern reveals the nature of the attack upon further investigation 
by the network operator. The range of source IP addresses specified in the pattern 
possibly contains botnet IP addresses. Attack flows are often reset when the load can
not be handled or the flows' states can not be recognized,
which indicates the attack is successfully launched. Packets with short intervals further provide 
direct support to this hypothesis. Unique properties of DDoS attack are indeed captured by 
this program!

Our next use case is based on Hulk, an attack similar to Slowloris. Hulk issues multiple 
HTTPS requests, trying to
keep them alive, adding more and more connections as time moves forward, and eventually 
overwhelming the webserver. Hulk requests have a high level of variety, 
adding difficulty to learning even with knowledge of the requests' contents. 
The synthesized NetQRE program to identify Hulk is as follows:
\begin{lstlisting}
( /_* A _*/  ( /_* B _*/ )*sum )max > 13
Where
A = [tcp.seq>=50%]
B = [tcp.fin==1]
\end{lstlisting}
The program first identifies a large sequence number, which is an indication 
that someone is trying to keep the connection long. This is followed by a large number 
of normally finished TCP connections. Connecting the two, it is not hard to guess 
someone is launching a long and slow attack. This is exactly how Hulk works to 
cause a DoS.

A takeaway in this use case is that \sys{} is able to build accurate classifiers 
without reliance on application-layer data, which is often encrypted. 
In some cases, even if application-layer data is desirable, \sys{} is able to build 
effective classifier simply by relying on features based on TCP and IP fields.

\noindent \textbf{Refinement by Human Knowledge}
Finally, an advantage of generating a program for classification is that it enables the operator 
to augment the generated NetQRE program with domain knowledge before deployment. 
For example, in the DDoS case, if they know that the victim service is purely 
based on TCP, they can append $[\mathit{ip.type} = TCP]$ to all predicates. Alternatively, 
if they know that the victim service is designed for 1000 requests per second, 
they can explicitly replace the arrival time interval with $1ms$. 
The modified program then is:
\begin{lstlisting}
( ( /_* A _* B _*/ )*sum /_* C _*/ )sum > 4
Where 
A = [ip.type = TCP]&&[ip.src_ip->[0%,50%]]
B = [ip.type = TCP]&&[tcp.rst==1]
C = [ip.type = TCP]&&[time_since_last_pkt<=1ms]
\end{lstlisting}


\subsection{Deployment Scenarios}
\label{sec:deploy}

We now describe three ways for network operators to deploy the output of \sys{}: (1) taking action hinted by the interpretation; (2) directly executing the NetQRE program as a monitoring system;  and (3) translating the NetQRE program to rules in other monitoring systems.

Revisiting the DDoS example in Section~\ref{sec:tuning}, in the first case, the operator may notice that the attack comes from a certain range of IP addresses. By further
investigation (for example by manually doing a binary search on the IP range in the NetQRE program), 
they find out that the attack is based on a cluster of machines with IP range $205.174.165.69 - 71$. Then this IP range can be blocked by a firewall to stop future attacks.


If the NetQRE program itself is to be used as a monitoring system, its runtime system can be directly 
deployed on any general purpose machine. Prior work~\cite{netqre} has shown that NetQRE generates performance that is comparable to optimized low-level implementations. Moreover, these programs can be easily compiled into other formats. In Appendix~\ref{appendix:B}, we demonstrate an example Bro~\cite{paxson1999bro} program translated from our synthesized NetQRE program for DDoS.


\subsection{Program Synthesis Performance}
\label{sec:synthesistime}

\begin{figure}[ht]
\centering
\includegraphics[width=0.45\textwidth]{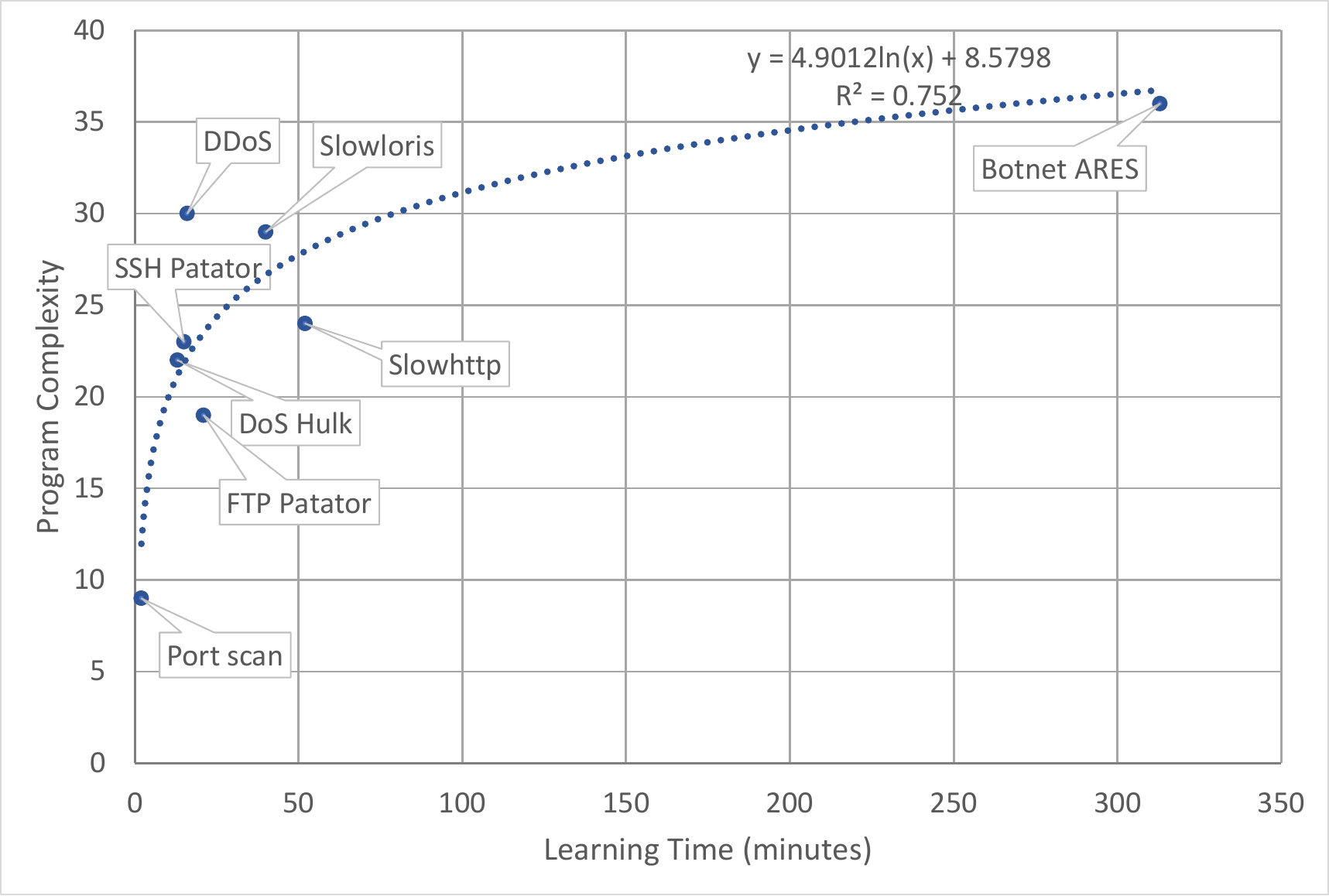}
\caption{Time-complexity relation}
\label{fig:time-comp}
\end{figure}

\noindent \textbf{Synthesis time:}
In our final experiment, we measure the performance of \sys{}, in terms of its program synthesis time. Our results show that \sys{} is able to identify candidate programs in reasonable time on complex real-world workloads. 

Figure~\ref{fig:time-comp} shows the program complexity (Y-axis) and synthesis (learning) time (in minutes). Program complexity is measured by the number of expansion
decisions (i.e. expansion of non-terminals in the syntax) during the search process. Not surprisingly, complex programs require more time to synthesize.
We further observe that \sys{} is able to synthesize complex programs with at least 20-30 terms. 
Despite the program complexity, synthesis time ranges from minutes to an hour, 
which is practical for many use cases and can be further reduced through parallelism. 

\begin{figure}[ht]
\centering
\includegraphics[width=0.45\textwidth]{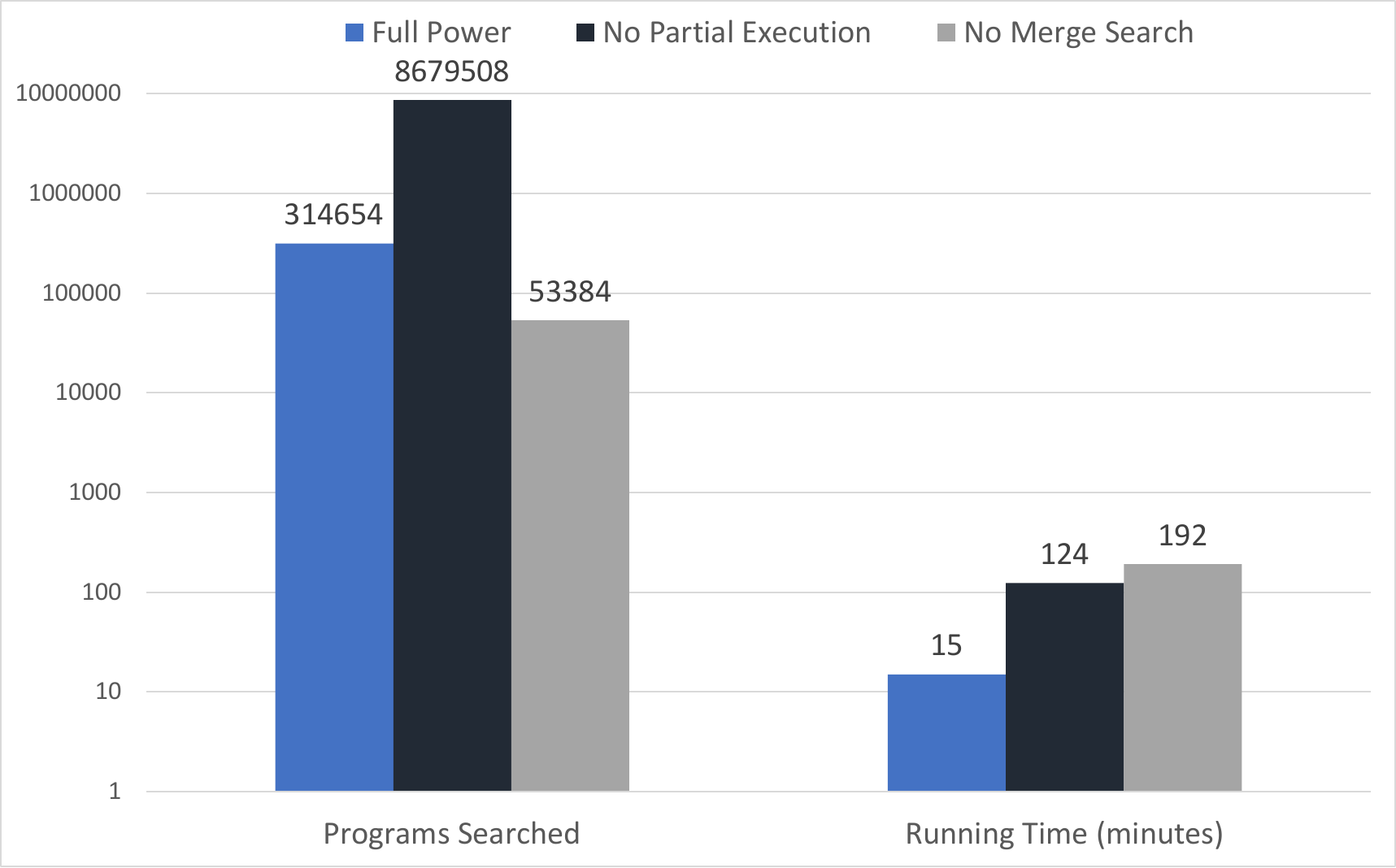}
\caption{Impact of optimizations on synthesis performance}
\label{fig:time-num}
\end{figure}

\begin{figure}[ht]
\centering
\includegraphics[width=0.35\textwidth]{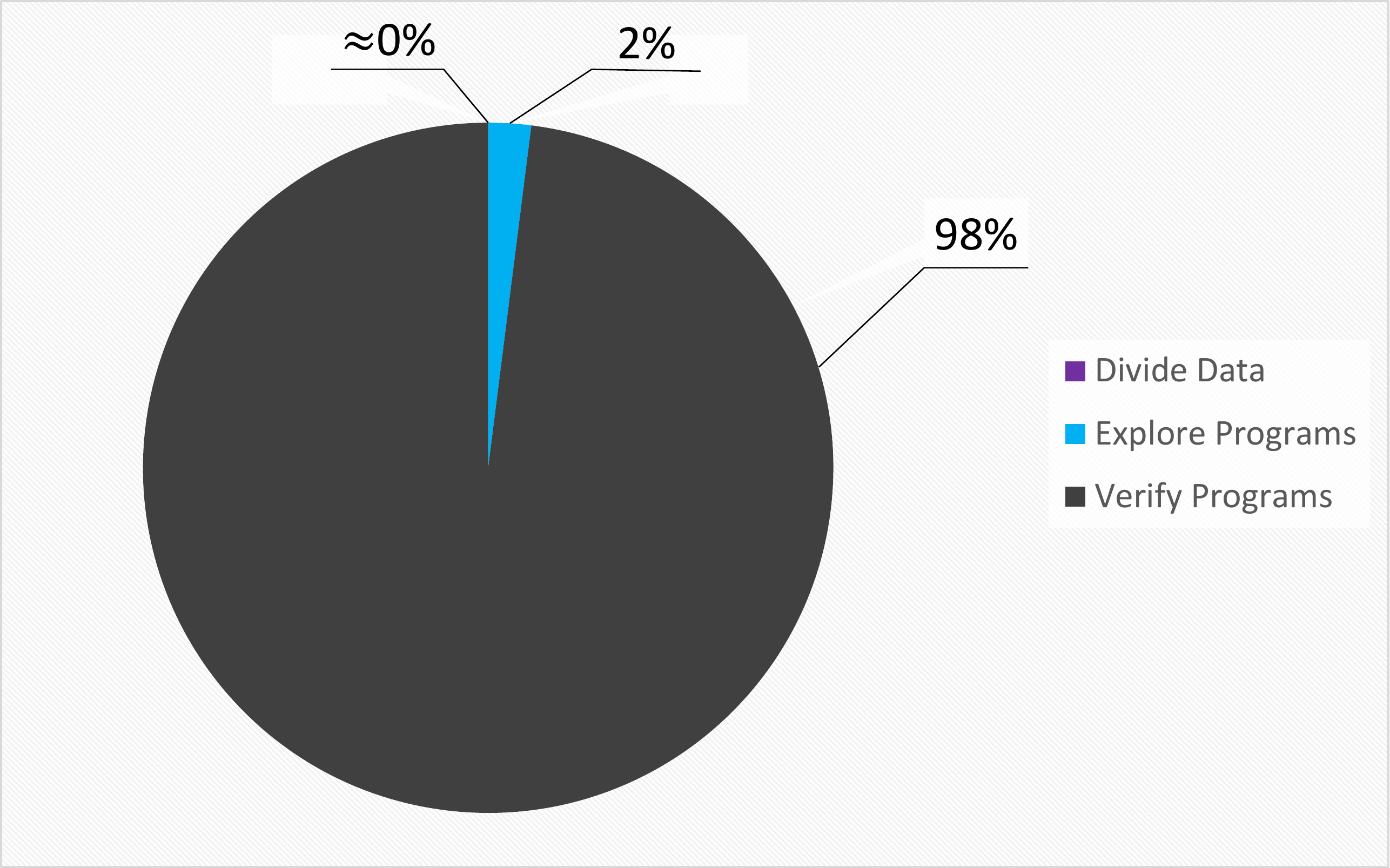}
\caption{Overhead of different steps}
\label{fig:profile}
\end{figure}

\noindent \textbf{Effectiveness of Optimizations. } We explore the effectiveness of the individual optimization strategies described in Section~\ref{sec:synthesis}.
In Figure~\ref{fig:time-num}, we compare the synthesis time and the number of programs searched for a fully optimized \sys{} against results from disabling each optimization. SSH Patator is used as the demonstrating example since it is moderately complex and can finish within reasonable time even after optimizations are disabled.

We observe that disabling partial execution optimization makes both programs searched and synthesis time significantly worse. Being able to prune early can indeed greatly reduce time wasted on unnecessary exploration and checking. By disabling merge search, although the number of programs searched decreases, the total synthesis time increases given the overhead of having to check each program against the entire data set. The synthesis can not finish within reasonable time if both are disabled.

Finally, to evaluate the parallelizability of the learning process, we profiled the task running entirely on
a single thread and show the time consumed by each phase in Figure~\ref{fig:profile}. Execution of candidate programs on examples takes up almost all execution time. The profiling result also shows that each individual  execution takes time in milliseconds (varying due to the size of the example). It supports that program synthesis is highly parallelizable in our architecture.

In summary, all optimization strategies are effective to speed up the synthesis process.
A synthesis task that is otherwise impossible to finish within practical time can now be done in less than 15 minutes.
Our architecture allows the parallelization of the most time-consuming part of the learning process.


\section{Related Work}


\noindent{\bf Automatic Generation of Network Configurations.} 
Broadly speaking, network traffic classification rule is a type of
network configuration. Apart from the forementioned competing systems
of \sys{}, there are also other lines of research that aim at
the automatic generation of different categories of network configurations.
EasyACL~\cite{liu2017automated} aims at synthesis of access control lists(ACL) from natural
language descriptions. Soumya et al~\cite{maity2012policy} instead derives ACL implementations
from network topology and input security policy specifications. 
NetGen~\cite{saha2015netgen}, NetComplete~\cite{el2018netcomplete} and Genesis~\cite{subramanian2017genesis} synthesize data plan routing
configurations based on SMT solvers given policy specifications
in regular expressions or customized policy languages. 
NetEgg~\cite{yuan2014netegg} instead takes examples provided by user to
generate routing configurations in an interactive way.
\sys{} focuses on network traffic classification and has a different 
target from them.

\noindent{\bf Unsupervised Learning Systems.} Unsupervised learning is useful for recognizing outliers and other types of "abnormal" flows~\cite{mishra2018detailed,zhang2015robust,xie2012subflow}, 
most notably in intrusion detection systems. Its ability to differentiate unknown types of traffic 
from the known cannot be replaced by \sys{}. 

The most notable shortcoming of unsupervised learning is relatively low accuracy and the difficulty
to create an inclusive set of "normal" traffic. \sys{} augments unsupervised learning systems by reducing the effort required for analyzing the reports. Traffic deemed abnormal can be fed into \sys{} to generate interpretable programs to speed up analysis.

\noindent {\bf Syntax-Guided Synthesis.} \sys{} builds on a large body of work on syntax-guided synthesis. However, synthesis techniques proposed
in this paper go beyond the state of the art, and have the potential to be applied to other applications of program synthesis. The partial execution technique is different from the classic compiler technique of \textit{partial evaluation}~\cite{futamura1999partial},
which aims to optimize a program for a faster but equivalent version by specialization.

Partial execution is similar to the idea in reference~\cite{lee2016synthesizing} (see also follow-ups 
~\cite{so2017synthesizing,so2018synthesizing}),
where the system learns plain regular expressions and overestimates the feasibility of
a non-terminal with a Kleene-star.
\sys{} generalizes it to the case when the program does classification based on numerical outputs.
Partial execution can possibly be applied to other synthesis tasks where the data is highly structured, and
its processing is tightly coupled with the language's syntax elements, e.g. learning
SQL expressions from examples~\cite{wang2017synthesizing}.

To the best of our knowledge, there is no prior work in program synthesis similar to our proposed merge search technique.
Merge search is not specific to \sys{}, and can be used in other synthesis tasks to allow the handling of large data sets. Finally, there is no prior work that solely uses program synthesis to perform accurate real-world large-scale classification. The closest work concerns simple low-accuracy programs synthesized as weak learners~\cite{cheung2012using}, and requires a separate SVM to assemble them into a classifier.


\section{Conclusion}

This paper presents \sys{}, which develops syntax-guided synthesis techniques to automatically 
generate NetQRE programs for classifying session-layer attack traffic. \sys{} can be used for generating 
network monitoring queries or signatures for intrusion detection systems from labeled traces.
Our results demonstrate three key value propositions for \sys{}, namely it requires minimal 
feature engineering for all use cases, is amenable to efficient implementation and compilation 
into code directly executable in legacy systems, and is easy to decipher and edit
to understand the nature of the attacks and adapt to customization needs.
While achieving these three benefits, \sys{} has accuracy comparable to state-of-the-art 
statistical and signature-based learning systems and requires synthesis time 
in minutes for most use cases.







\bibliographystyle{plain}

\appendix

\section{NetQRE Examples}
\label{appendix:A}


Below is the full list of learnt classification programs on CICIDS2017
database. The medium of the recommended ranged is taken as the default
threshold.
\begin{lstlisting}
Slowloris:  
( / _* A _* / (/ _* B _* /)*sum )sum > 7
Where
A = [ip.src_ip->[50%,100%]]
B = [ip.src_ip->[50%,62.5%]]&&[tcp.syn==1]

Slowhttp:   
( / _* A _* / ( / _* B _* / )*sum )max > 57
Where
A = [tcp.ack==0]
B = [ip.src_ip->[25%,50%]]

DoS Hulk:   
( / _* A _* / ( / _* B _* / )*sum )max > 13
Where
A = [tcp.seq>=50%]
B = [tcp.fin==1]

SSH Patator:
( ( / _* A _* / )*sum / _* B _* / )max > 109
Where
A = [tcp.psh==1]
B = [tcp.win<=50%]

FTP Patator:
( / _* _ / ( / _* A _* / )*sum )max > 98
Where
A = [tcp.src_port->[25%,50%]]

Botnet ARES:
( ( / _* A _* B _* C _* / )*sum / _* D _* / )sum > 9
Where
A = [tcp.fin==1]
B = [tcp.syn==1]
C = [ip.dst_ip->[50%,100%]]
D = [ip.len->[0%,50%]]

DDoS:       
( ( /_* A _* B _*/ )*sum /_* C _*/ )sum > 4
Where 
A = [ip.src_ip->[0%,50%]]
B = [tcp.rst==1]
C = [time_since_last_pkt<=50%]

Port Scan:  
( ( / _* _ / )*max )sum|tcp.dst_port > 9
\end{lstlisting}

\smallskip

\section{Translated Bro Program}
\label{appendix:B}

We show below a NetQRE program translated into Bro for detecting DDoS based on 
the generated DDoS program in Appendix~\ref{appendix:A}. The events can be added by plugins.

\begin{lstlisting}
type StateType: enum {Init, IPMatched, RSTMatched};
global st: StateType = Init;
global counter = 0;
global timestamp: Time = CurrentTime;

function initialize() {
    st = Init;
    counter = 0;
    timestamp = CurrentTime;
}
event tcp_init(src_ip: IPAddress, ......) {
    if (CurrentTime - timestamp > Timeout) {
        initialize;
    }
    if (src_ip in SuspectRange && st == Init) {
        st = IPMatched;
    }
}
event tcp_reset(......) {
    if (CurrentTime - timestamp > Timeout) {
        initialize;
    }
    if (st == IPMatched) {
        counter += 1;
        st = RSTMatched;
    }
}
event short_interval(interval: Time, ......) {
    if (CurrentTime - timestamp > Timeout) {
        initialize;
    }
    if (interval < Threshold) {
        counter += 1;
    }
    if (counter > 4) {
        Notice("DDoS!");
        initialize;
    }
}
\end{lstlisting}

\end{document}